\documentclass[amsmath,amsfonts,prb,twocolumn,longbibliography]{revtex4-1}
\usepackage{graphicx}
\usepackage{epsfig}
\usepackage{bm}
\usepackage{bbm}
\usepackage{dcolumn}
\usepackage{amsmath}
\usepackage{amssymb}
\usepackage{amsfonts}
\usepackage{color}
\usepackage{soul}

\newcommand{\beg}{\begin{equation}}
\newcommand{\en}{\end{equation}}
\newcommand{\veps}{\varepsilon}
\newcommand{\eps}{\epsilon}

\newcommand{\bk}{\mathbf k}
\newcommand{\br}{\mathbf r}

\newcommand{\bx}{\mathbf x}
\newcommand{\dg}{^\dagger}

\begin{document}

\title{Magnetic impurities in a strongly coupled superconductor}

\author{Samuel Awelewa}
\address{Department of Physics, Kent State University, Kent, OH 44242, USA}

\author{Maxim Dzero}
\address{Department of Physics, Kent State University, Kent, OH 44242, USA}

\begin{abstract}
We revisit certain aspects of a problem concerning the influence of carrier scattering induced by magnetic impurities in metals on their superconducting properties. Superconductivity is assumed to be driven by strong electron-phonon interaction. We use the self-consistent solution of the Nagaoka equations for the scattering matrix together with the Migdal-Eliashberg theory of superconductivity to compute the energy of the in-gap bound states, superconducting critical temperature and tunneling density of states for a wide range of values of the Kondo temperature and impurity concentrations. It is found that similar to the case of the weak coupling (BCS) superconductors there is only one pair of the bound states inside the gap as well as re-entrant superconductivity for the case of antiferromagnetic exchange coupling between the conduction electrons and magnetic impurities. In agreement with the earlier studies we find that the gapless superconductivity can be realized which in the case of antiferromagnetic exchange requires much smaller impurity concentration. Surprisingly, in contrast with the weakly coupled superconductors we find that superconducting transition exhibits two critical temperatures for the ferromagnetic exchange coupling. 
\end{abstract}


\date{\today}

\maketitle

\section{Introduction}
A problem of an interplay between magnetism, appearing due to either itinerant or localized degrees of freedom, and superconductivity still remains one of the most fascinating problems in condensed matter physics. \cite{UBe13,Ce122,Hewson1993,Coleman2007} Despite its long history which dates back to observation of superfluidity in $^3$He and the experiments by Berndt Matthias et al. \cite{Volovik2009,Matthias1963}, the last several decades of research lead to the development of novel physical concepts such as quantum criticality \cite{Sachdev2011,QCP-Review,Qi-RMP20} as well as fascinating physical effects such as topological superconductivity, \cite{UTe2Dan,Aoki2022,UTe2Ref2,UTe2Ref3,UTe2} unconventional Fulde-Ferrell-Larkin-Ovchinnikov superconductivity \cite{FFLO2024} and thermodynamic anomalies in iron-based and cuprate superconductors.\cite{Putzke,Joshi,Auslaender,Hashimoto1,Hashimoto2,Huang,Carvalho,Khodas,Levchenko,Chowdhury1,Chowdhury2} This problem has received a renewed interest and is still being actively explored in the context of terahertz experiments in disordered superconductors \cite{Jujo2018,Higgs2024,LinResponse2024} and iron-based superconductors.\cite{ChubukovAnRev2012,QSi2023,Dzero}

Perhaps the most basic aspect of that problem concerns the fate of a paramagnetic impurities embedded in the bulk of a conventional superconductor. Thanks to the seminal work by Abrikosov and Gor'kov \cite{AG1961} it has been known that  paramagnetic impurities lead to a strong suppression of superconductivity in Bardeen-Cooper-Schrieffer (BCS) superconductors where the interaction between electron and lattice degrees of freedom is assumed to be weak.\cite{bcstheory} It was later discovered, that the presence of the magnetic impurities also leads to the appearance of the impurity bound states \cite{Yu,Shiba,Rusinov,Balatsky-RMP} as well as the spin-glass state co-existing with superconductivity. \cite{VityaG-2002} 

The earlier theories considered the case of the ferromagnetic exchange between the spin of conduction electrons and paramagnetic moments on impurities and, as such, ruled out the possibility for the development of the Kondo effect - the experimental observation of the resistivity minimum with decreasing temperatures in metallic alloys with paramagnetic impurities \cite{Kondo64,Nagaoka1965,MHZNormal} - since its microscopic origin was not understood at that time.\cite{Abrikosov1988} Consequently, it took another decade between the understanding of the microscopic origin of the Kondo effect \cite{Kondo64} and the appearance of the first theory which addressed how the quantum dynamics associated with the scattering of electrons on magnetic impurity may affect superconductivity.\cite{MHZ1970I} Specifically, these works focused on understanding how suppression of superconductivity takes place when the single impurity Kondo temperature and superconducting critical temperature become comparable to each other,  $T_c\sim T_K$.\cite{MHZ1970I,MHZ1971,MHZ1970II,Schuh1978,MIN} In particular, it was found that when $T_c$ is suppressed far below $T_K$, $T_c$ never goes to zero and the superconductivity persists due to the screening of the magnetic moment on the impurity by conduction electrons leading the characteristic re-entrant shape of the superconducting critical temperature as a function of impurity concentration. \cite{Schuh1978,MIN} It must be emphasized that even though the temperature range in which the results of the above mentioned theories are valid is limited to the temperatures not too far below the Kondo temperature and still assumes that concentration of impurities is small enough to prevent the system from developing spatial inhomogeneities,  these theoretical results have been confirmed experimentally in a series of independent experiments. \cite{Maple1,Maple2,Maple3,Maple4} Furthermore, the methodology developed in Refs. \cite{Nagaoka1965,MHZNormal} have been successfully used by one of us to explain the unusual transport properties in so-called charge-Kondo systems.\cite{DzeroCK,IanFisher2005}

It is worth pointing out here that the results of the earlier theories \cite{MHZ1970I,MHZ1971,MHZ1970II,Schuh1978,MIN} have also been extended to the case of the dense paramagnetic alloys by Barzykin and Gor'kov who demonstrated that phonon-mediated superconductivity may survive even in the dense Kondo lattice-like environment \cite{LPG2005}. Similarly,  we have recently considered a problem of competition between the onset of the coherence and phonon mediated strong-coupling superconductivity in the Kondo lattice and found that superconductivity precludes the emergence of the coherent heavy-fermion state when the system is in the local moment regime \cite{Awelewa2024}.

In the papers mentioned above superconductivity has been described using the formalism of the weak coupling BCS theory. To the best of our knowledge, Schachinger, Daams and Carbotte \cite{Schach1980} were first who addressed the problem of the paramagnetic impurities in superconductors without relying on an assumption of weak electron-phonon coupling. The authors of Refs. \cite{Schach1980,Schachinger1985} have employed the methodology developed in Refs.\cite{MHZ1970I,MHZ1971,MHZ1970II} to find the corrections to the Eliashberg equations due to scattering on the paramagnetic impurities.  In these papers \cite{Schach1980,Schachinger1985} the dependence of the superconducting critical temperature of a strong coupling superconductor was computed and it was found that the effect of Kondo impurities was overestimated by a factor of $1/(1+\lambda)$ where $\lambda$ is the dimensionless electron-phonon coupling constant. For example, in order to obtain good agreement with the experimental data one had to re-scale the value of the single-ion Kondo temperature $T_K$ by a factor of $\exp[2.56/(1+\lambda)]$. 

It is important for us to mention, however, that the derivation of the expressions for the critical temperature as well as tunneling characteristics have relied on several significant simplifications. \cite{Schach1980,Schachinger1985} For example, the frequency dependence of the scattering matrix which accounts for the Kondo resonance has been neglected. On the other hand, given the fact that both Eliashberg equations as well as equation for the scattering matrix need to be analyzed numerically due to their inherent nonlinearity, it seems to us that these approximations are not necessary except in the case when one needs to make simple estimates to ensure that the results of the numerical solution are indeed correct. It thus remains not entirely clear how these approximation will affect the dependence of the critical temperature on the impurity concentration, for example. It must also be mentioned that the approach taken in \cite{Schach1982} have considered the case of ferromagnetic exchange coupling thus completely ignoring an interplay between the Kondo effect and strong coupling superconductivity and focusing mostly on the effects of strong electron-phonon coupling on the in-gap bound states.  

In view of the discussion above, in this paper we revisit the problem of a paramagnetic impurities in conventional superconductors with strong electron-phonon interactions for the arbitrary sign of the exchange coupling between spin of a conduction electron and magnetic moment of an impurity. We compute the energy of the bound states, critical temperature and single particle density of states as a function of the exchange coupling and impurity concentration. Our analysis is based on iterative self-consistent solution of the Eliashberg equations together with the equation for the scattering matrix without relying on any approximations for the scattering matrix. \cite{MHZ1970I,MHZ1971,MHZ1970II} Our main finding is that in a strongly coupled superconductor, paramagnetic impurities substantially affect the superconducting critical temperature. Specifically, in a case of ferromagnetic coupling we find that there are two critical temperatures $T_{\textrm{c1}}$ and $T_{\textrm{c2}}$, so that superconductivity exists for temperatures $T_{\textrm{c1}}\leq T\leq T_{\textrm{c2}}$, when the impurity concentration does not exceed a certain critical value $c_{\textrm{imp}}^*$. Subsequently, no superconductivity exists for $c_{\textrm{imp}}\geq c_{\textrm{imp}}^*$. To the best of our knowledge this is new result.  In the opposite case of the antiferromagnetic coupling when the single ion Kondo temperature is smaller than the superconducting critical temperature for a clean superconductor, $T_K<T_{\mathrm{c0}}$, the dependence of the critical temperature on the concentration of impurities is similar to what one finds for the case of the BCS superconductor displaying the characteristic re-entrant shape provided the value of the exchange coupling is not too large. 

\section{Model and basic equations}
We consider a model which describes conduction electrons which strongly interact with dispersionless phonons (Holstein model) in the presence of magnetic impurities. The exchange interaction between the conduction electrons and magnetic impurities is assumed to be arbitrary.  The model Hamiltonian is
\beg\label{Eq1}
\begin{split}
\hat{\cal H}&=\sum\limits_{ij\sigma}h_{ij}\hat{c}_{i\sigma}\dg\hat{c}_{j\sigma}+\sum\limits_{i}\left[\frac{\hat{P}_i^2}{2M}+\frac{M\Omega^2\hat{x}_i^2}{2}\right]\\&+\alpha\sum\limits_{i\sigma}\hat{x}_i\hat{c}_{i\sigma}\dg\hat{c}_{i\sigma}-J\sum\limits_{\bk\bk'}\sum\limits_{\alpha\beta}
\left({\vec S}\cdot{\vec \sigma}\right)_{\alpha\beta}\hat{c}_{\bk\alpha}\dg\hat{c}_{\bk'\beta}.
\end{split}
\en
Here $h_{ij}$ is the hopping matrix elements for the conduction electrons, $i,j$ label the lattice sites, $\hat{c}_{i\sigma}$, $\hat{c}_{i\sigma}\dg$ are the fermionic annihilation and creation operators, $\sigma=\pm 1/2$ is the spin projection, $\hat{P}_i$, $\hat{x}_i$ are ion conjugate momentum and position operators, $M$ is the mass of an ion, $\Omega$ is the frequency of the optical mode, $\alpha$ is the coupling of the electrons to the ionic displacement, ${\vec S}$ is an impurity spin operator and $J$ is the magnetic exchange coupling. We will assume that the concentration of impurities is very small, so that the interactions between the impurities can be fully ignored, i.e. we assume that for the case of antiferromagnetic exchange $J<0$, single-ion Kondo temperature far exceeds the characteristic temperature scale for the spin-spin exchange interaction, $T_K\gg T_{\textrm{RKKY}}$. 

It can be shown \cite{Chubukov2020,EmilEli2,Emil2024} that the model Hamiltonian (\ref{Eq1}) for $J=0$ can be used to describe the emergence of superconductivity provided that the effective coupling constant is not too large.  As we have already mentioned above it is our goal to investigate how the onset of this instability is affected by the scattering on magnetic impurities described by the last term in (\ref{Eq1}). 
The central quantity to our analysis will be the single particle propagator $\hat{G}(x,x')$  $(x=(\bx,\tau))$ in the imaginary time representation \cite{AGD}, 
which is a matrix in Nambu space:
\beg\label{Gxxp}
{G}_{\mathrm{ab}}(x,x')=\langle\hat{T}_\tau\{{\Psi}_{\mathrm{a}}(x)\Psi_{\mathrm{b}}\dg(x')\}\rangle,
\en
where $\hat{\Psi}\dg(x)=[\psi_{\uparrow}\dg(x), ~\psi_{\downarrow}(x)]$ is the Nambu spinor. Next, it will be convenient to consider (\ref{Gxxp}) in the momentum and Matsubara frequency representation $\hat{G}_{\bk\bk'}(i\omega_n)$, where $\omega_n=\pi T(2n+1)$ and $T$ is temperature. Then the scattering matrix $\hat{t}(i\omega_n)$ in Nambu space can be defined according to \cite{MHZ1971,Hewson1993}
\beg\label{twn}
\begin{split}
\hat{G}_{\bk\bk'}(i\omega_n)&=\hat{\cal G}_\bk(i\omega_n)\delta_{\bk\bk'}\\&+c_{\textrm{imp}}J\hat{\cal G}_\bk(i\omega_n)\hat{t}(i\omega_n)\hat{\cal G}_{\bk'}(i\omega_n).
\end{split}
\en
Here $c_{\textrm{imp}}$ determines the concentration of magnetic impurities, $\hat{\cal G}_{\bk}(i\omega_n)$ is the single-particle propagator in a clean superconductor. Lastly, the second term in (\ref{twn}) includes the frequency-dependent scattering matrix due to the scattering on a single impurity which  has the following matrix form \cite{MHZ1970I}
\beg\label{tiwn}
\hat{t}(i\omega_n)=\left[\begin{matrix} t_1(i\omega_n) & t_2(i\omega_n) \\ t_2(i\omega_n) & t_1(i\omega_n) \end{matrix}\right].
\en
Matrix function (\ref{Gxxp}) satisfies the Dyson equation \cite{Kamenev2011}
\beg\label{Dyson}
\left(\hat{\cal G}_0^{-1}-\hat{\Sigma}\right)\circ\hat{G}(x,x')=\hat{\mathbbm{1}},
\en
where $\hat{\cal G}_0^{-1}=i\partial_t\hat{\tau}_3-\hat{\cal H}_0$, $\hat{\cal H}_0$ contains non-interacting terms of the Hamiltonian (\ref{Eq1}),    
$\hat{\tau}_3$ is the third Pauli matrix which acts in the Nambu space, 
$x=(\br,\tau)$, $\hat{\Sigma}(\br,\tau)$ is the self-energy part which is given by the sum of the corresponding self-energy contributions from the electron-phonon interaction and impurity scattering,  $\hat{\Sigma}=\hat{\Sigma}_{\textrm{e-ph}}+\hat{\Sigma}_{\mathrm{imp}}$. Given expression (\ref{Gxxp}), for the self-energy part we can write down the following equation 
\beg\label{Self}
\begin{split}
\hat{\Sigma}(i\omega_n)&=T\sum\limits_{i\omega_l}{\cal D}(i\omega_n-i\omega_l)\int\frac{d^3\bk}{(2\pi)^3}\hat{\tau}_3\hat{G}_{\bk\bk}(i\omega_l)\hat{\tau}_3\\&+c_{\mathrm{imp}}J\hat{t}(i\omega_n).
\end{split}
\en
In this equation  ${\cal D}(i\eps_l)$ is the phonon propagator which without loss of generality we choose in the form corresponding to the Holstein model 
${\cal D}(i\eps_l)=\nu_F^{-1}\lambda\Omega^2/(\eps_l^2+\Omega^2)$ where $\lambda$ is the dimensionless electron-phonon coupling constant,  $\nu_F$ is the single particle density of states at the Fermi level, $\Omega$ is the frequency of the optical phonon and $\eps_l=2\pi Tl$ are the bosonic Matsubara frequencies.\cite{EmilEli1} It is worth noting that the correlator $\hat{\cal G}_\bk(i\omega_n)$ can be cast into the following form 
\beg\label{calGk}
\hat{\cal G}_\bk^{-1}(i\omega_n)=i\omega_n\hat{\tau}_0-\xi_\bk\hat{\tau}_3-\hat{\Sigma}_{\textrm{e-ph}}(i\omega_n), 
\en
where $\xi_\bk$ is the energy of a single particle with momentum $\bk$, $\hat{\tau}_{0,1}$ are the unit and the first Pauli matrix correspondingly. Furthermore, assuming the particle-hole symmetry, for the self-energy part $\hat{\Sigma}_{\textrm{e-ph}}(i\omega_n)$ one finds the following expression 
\beg\label{Sigman}
\hat{\Sigma}_{\textrm{e-ph}}(i\omega_n)=i\omega_n[1-Z(i\omega_n)]\hat{\tau}_0+\Phi(i\omega_n)\hat{\tau}_1.
\en
In this expression function $\Phi(i\omega_n)$ accounts for the onset of the Cooper pairing, while function $Z(i\omega_n)$ describes the self-energy re-normalization due to the interaction between the electrons and phonons. 
Expression (\ref{Sigman}) together with (\ref{tiwn}) suggest that we can adopt the similar ansatz for the self-energy part in (\ref{Self}). As a result we derive the following system of equations for the functions $Z(i\omega_n)$ and $\Delta_n=\Phi(i\omega_n)/Z(i\omega_n)$:
\beg\label{ModifyEli}
\begin{split}
\omega_n[Z(i\omega_n)-1]&=\pi T\sum\limits_{\omega_m}\lambda_{nm}\frac{\left[\omega_m-\Delta_m{\cal T}(i\omega_m)\right]}{\sqrt{\omega_m^2+\Delta_m^2}}\\&-ic_{\textrm{imp}}Jt_1(i\omega_n), \\
\Delta_nZ(i\omega_n)&=\pi T\sum\limits_{\omega_m}\lambda_{nm}\frac{\left[\Delta_m+\omega_m{\cal T}(i\omega_m)\right]}{\sqrt{\omega_m^2+\Delta_m^2}}\\&+c_{\textrm{imp}}Jt_2(i\omega_n).
\end{split}
\en
Here we use the shorthand notation $\lambda_{nm}={\cal D}(i\omega_n-i\omega_m)$, we introduced  function
\beg\label{CalT}
{\cal T}(i\omega_n)=c_{\textrm{imp}}J\frac{\left[{\omega_nt_2(i\omega_n)}-{i\Delta_nt_1(i\omega_n)}\right]}{(\omega_n^2+\Delta_n^2)Z(i\omega_n)},
\en
which is an odd function of Matsubara frequencies, ${\cal T}(-i\omega_n)=-{\cal T}(i\omega_n)$. Note that (i) the impurity scattering corrections described by  ${\cal T}(i\omega_n)$ enter into the Eliashberg equations  (\ref{ModifyEli}) with the opposite sign as it should be expected for scattering on magnetic impurities\cite{AG1961} and (ii) the combination of terms appearing in the second equation under the Matsubara sum coincides with the corresponding self-consistency equation in the weak coupling theory.\cite{MHZ1970I} We also note that for $J=0$ equations (\ref{ModifyEli}) acquire the well-known form of the Eliashberg equations (see e.g. Ref. \onlinecite{EmilEli1}). 

\subsection{Equation for the scattering matrix}
At the level of the saddle-point approximation, equations (\ref{ModifyEli},\ref{CalT}) describe how superconductivity is modified by the presence of the magnetic impurities. The analysis of these equations, however, requires the knowledge of the normal and anomalous components of the impurity scattering matrix, $t_1(i\omega_n)$ and $t_2(i\omega_n)$. One way to compute these functions is to use the method of the equations of motion for the single particle correlations, which in the context of the Kondo impurity problem was pioneered by Nagaoka \cite{Nagaoka1965} and then further developed by M\"{u}ller-Hartmann and Zittartz.\cite{MHZNormal,MHZ1970I,MHZ1971,MHZ1970II} In this formalism the central role is played by the matrix function $\hat{F}(i\omega_n)$:
\beg\label{hatFz}
\hat{F}(i\omega_n)={J}\sum_\bk\hat{\cal G}_\bk(i\omega_n)=\left[\begin{matrix} F_1(i\omega_n) & F_2(i\omega_n) \\ 
F_2(i\omega_n) & F_1(i\omega_n)\end{matrix}\right]
\en
with $\hat{\cal G}_\bk(i\omega_n)$ given by expression (\ref{calGk}).
Following the avenue of Refs. \cite{MHZNormal,MHZ1970I} we obtain the following self-consistent equation for the scattering matrix $\hat{t}(i\omega_n)$:
\beg\label{Eq4tiwn}
\begin{split}
&\left[1-\frac{S(S+1)}{4}\hat{F}^2(z)+\hat{\cal S}_{\omega_n}\left\{\frac{\hat{F}(i\omega_n)-\hat{F}(z)}{z-i\omega_n}\right.\right.\\&\left.\left.\times
\left[1+\left(\hat{F}(i\omega_n)-\hat{F}(z)\right)\hat{t}(i\omega_n)\right]\right\}\right]\hat{t}(z)\\&=\frac{S(S+1)}{4}\hat{F}(z)+\hat{\cal S}_{\omega_n}\left\{\frac{\hat{F}(i\omega_n)-\hat{F}(z)}{z-i\omega_n}\hat{t}(i\omega_n)\right\}.
\end{split}
\en
In this equation we introduced $\hat{\cal S}_{\omega_n}\left\{g_{\omega_n}\right\}\equiv T\sum\limits_{\omega_n}e^{i\omega_n0+}g_{\omega_n}$, $T$ is temperature and $S$ is the magnitude of the impurity spin. Matrix equation (\ref{Eq4tiwn}) can be further simplified by using the following matrix identity:
\beg\label{MainIdentity}
\left(\begin{matrix} a & b \\ b & a
\end{matrix}\right)\left(\begin{matrix} 1 & -1 \\ -1 & 1
\end{matrix}\right)=(a-b)\left(\begin{matrix} 1 & -1 \\ -1 & 1
\end{matrix}\right).
\en
Thus, with the help (\ref{MainIdentity}) equation (\ref{Eq4tiwn}) reduces to an algebraic equation for the function $t_{-}(i\omega_n)=t_1(i\omega_n)-t_2(i\omega_n)$ in which one needs to replace $\hat{F}$ with $F_{-}(i\omega_n)=F_1(i\omega_n)-F_2(i\omega_n)$. For a given set of $Z(i\omega_n)$ and $\Delta_n$ the resulting equation for $t_{-}(i\omega_n)$ can be easily solved by iterations. Consequently, taking into account that $t_1(-z)=-t_1(z)$
and $t_2(-z)=t_2(z)$ as well as that $\omega_{n<0}=-\omega_{-n-1}$, one finds $t_1(z)=(1/2)[t(z)-t(-z)]$, $t_2(z)=-(1/2)[t(z)+t(-z)]$.

%%%%%%%%%%%%% Fig1: Tc AFM %%%%%%%%%%%%%%%%%
\begin{figure}
\includegraphics[width=0.90\linewidth]{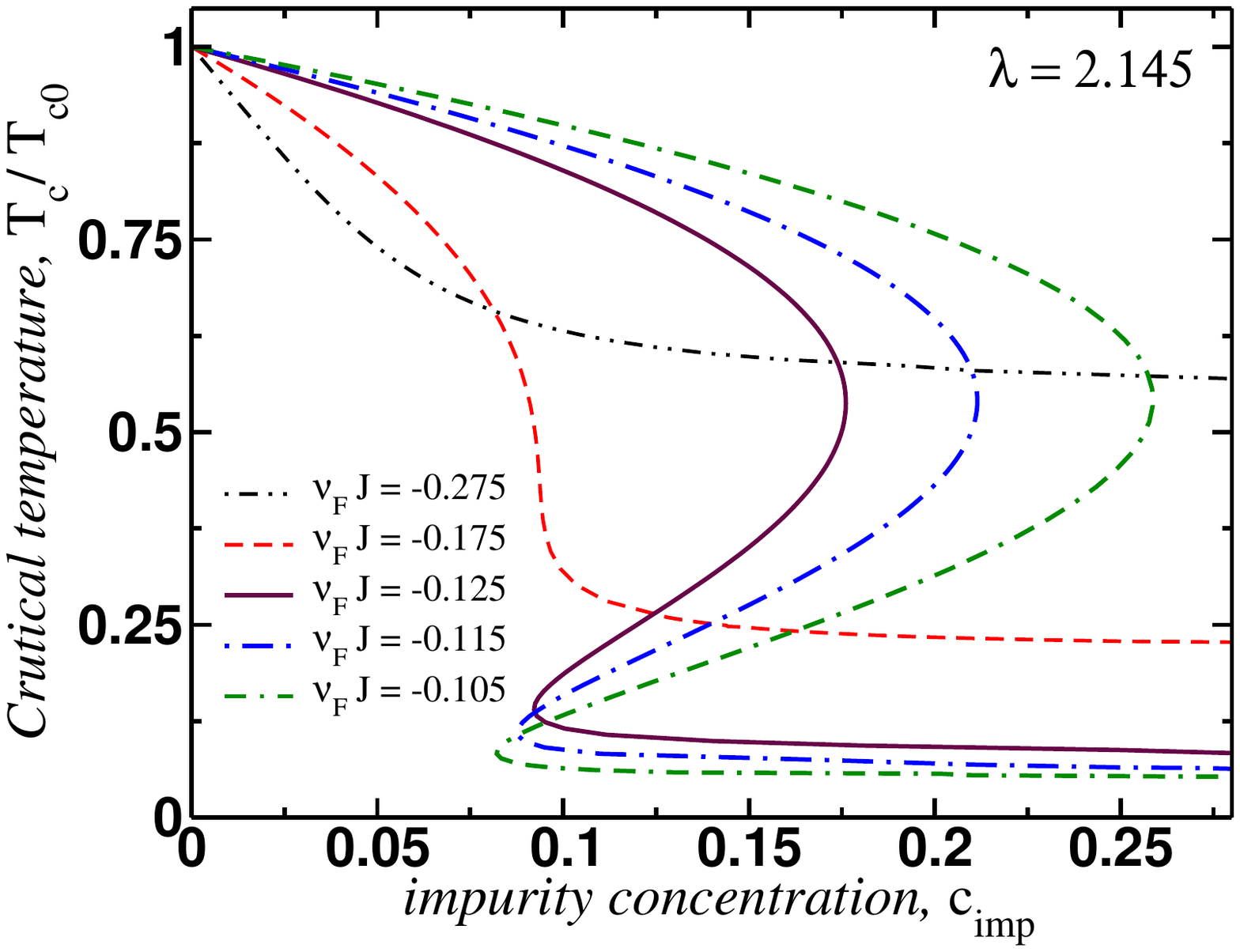}
\caption{Dependence of the superconducting critical temperature on the concentration of magnetic impurities for various values of the dimensionless antiferromagnetic exchange coupling computed from equation (\ref{Eq4PhinTc}) for $S=1/2$. The calculation of the Matsubara summations have been limited to the range $m\in[-N-1,N]$ with $N=512$.}
\label{Fig1-AFM-Tc}
\end{figure}
%%%%%%%%%%%%%%%%%%%%%%%%%%%%%%%%%%%%%%%%%%%%%%%

\section{Critical temperature of the superconducting transition}
\begin{figure}
\includegraphics[width=0.90\linewidth]{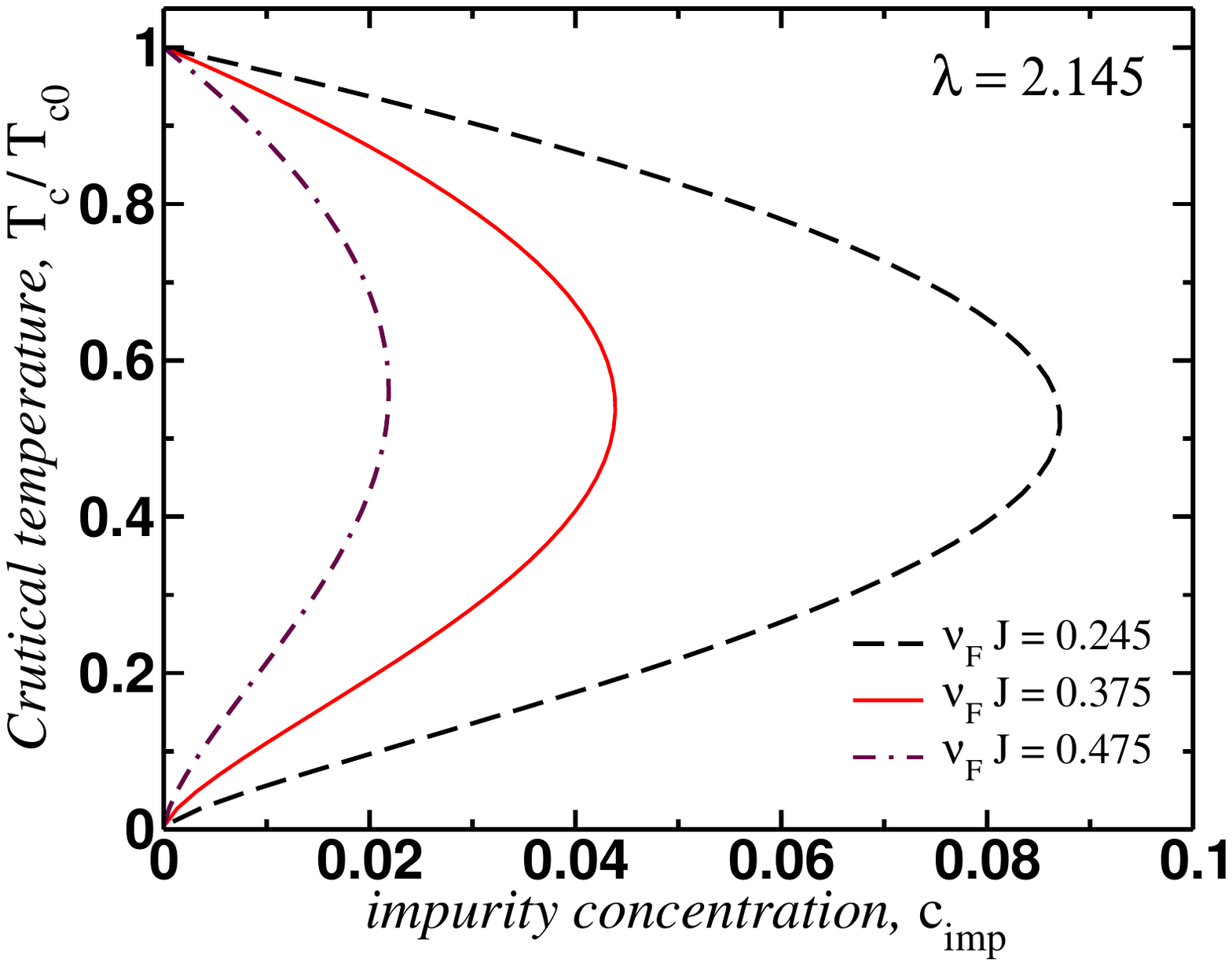}
\caption{Dependence of the superconducting critical temperature on the concentration of magnetic impurities for various values of the dimensionless ferromagnetic exchange coupling computed from equation (\ref{Eq4PhinTc}) for $S=1/2$. The calculation of the Matsubara summations have been limited to the range $m\in[-N-1,N]$ with $N=512$.}
\label{Fig2-FM-Tc}
\end{figure}

In order to compute the dependence of the critical temperature of the superconducting transition as a function of impurity concentration, we can set 
$\Delta_n=0$ in the first equation (\ref{ModifyEli}). Consequently, in the second equation in (\ref{ModifyEli}) we need to retain all the contributions which are linear in $\Phi(i\omega_n)=\Delta_nZ(i\omega_n)$. Since that equation involves the anomalous part of the scattering matrix $t_2(i\omega_n)$ we need to determine the coefficients of the linear expansion 
\beg\label{LinExpand}
t_2(i\omega_n)=\sum\limits_{m=-\infty}^\infty \Gamma_{nm}\Phi(i\omega_m).
\en
This task can be accomplished by expanding each term in the equation for the scattering matrix (\ref{Eq4tiwn}) up to the first order in powers of $\Phi(i\omega_n)$. The details of the calculation are given in the Appendix \ref{lineqt2}. From the numerical computation we found that the linear expansion coefficients $\Gamma_{nm}$ are sharply peaked for $n=m$, i.e. $|\Gamma_{nn}|\gg|\Gamma_{nm}|$ which allows us to approximate $t_2(i\omega_n)\approx \Gamma_{n}\Phi(i\omega_n)$. Thus, the final form of the equation for the pairing field $\Phi(i\omega_n)\equiv\Phi_n$ reads
\beg\label{Eq4PhinTc}
\begin{split}
&\Phi_n[1+c_{\textrm{imp}}J\Gamma_n]=\pi T_c\sum\limits_m\frac{\lambda_{nm}{\Phi}_m}{|\tilde{\omega}_m|}\\&\times\left\{1-c_{\textrm{imp}}J\left[\Gamma_m-\frac{it_1(i\omega_m)}{\tilde{\omega}_m}\right]\right\}.
\end{split}
\en
Clearly, this equation has a form of an eigenvalue equation $\eta\Phi_n=\sum\limits_ma_{nm}\Phi_m$ and the value of the critical temperature will be determined from the condition that the smallest eigenvalue equals one, $\eta=1$. In Figs. \ref{Fig1-AFM-Tc},\ref{Fig2-FM-Tc} we present the dependence of the critical temperature on the concentration of magnetic impurities for a fixed value of the electron-phonon coupling and for varying strength of the dimensionless exchange coupling $\nu_FJ<0$ and $\nu_FJ>0$ correspondingly. 
%%%%%%%%%%%%% Fig2: Tc FM %%%%%%%%%%%%%%%%%
%%%%%%%%%%%%%%%%%%%%%%%%%%%%%%%%%%%%%%%%%%%%%%%

In Fig. \ref{Fig1-AFM-Tc} which shows the suppression of superconductivity for the antiferromagnetic exchange coupling, we can clear distinguish 
three different regimes. The first one corresponds to a case when the single impurity Kondo temperature $T_K$ exceeds the critical temperature $T_K\gg T_{\textrm{c0}}$. This regime describes the situation when the impurity moment as almost fully screened by conduction electrons and as a result the suppression is very weak. The second regime accounts for the case when $T_K\sim T_{\textrm{c0}}$: initially strong suppression of superconductivity is succeeded by slow decrease in $T_c$. Lastly, the third regime corresponds to the case when $T_K<T_{\textrm{c0}}$ when $T_c(c_{\textrm{imp}})$ has a characteristic inverted $S$-shape describing the effect of re-entrant superconductivity, \cite{MHZ1971} which has been dubbed in the literature as 'three-$T_c$' behavior. \cite{Maple3,Maple4} Compared to the case of the weak coupling superconductivity, the new feature here is a crossover from the 'three-$T_c$' to a single-$T_c$ behavior as the value of the Kondo temperature increases. It is worth mentioning that our results in Fig. \ref{Fig1-AFM-Tc} agree qualitatively with those reported in Ref. \cite{Schachinger1985}

In Fig. \ref{Fig2-FM-Tc} we present the dependence of the critical temperature on the impurity concentration for the case of the ferromagnetic exchange. Here we observe quite drastic differences compared to the results of the weak-coupling theory and, at the same time, similarity with the dependence of the critical temperature on the size of the hybridization in the Kondo lattice. Specifically, we find that depending on the strength of the exchange coupling, the superconductivity is always limited to a finite temperature region $T_{\textrm{c1}}\leq T\leq T_{\textrm{c2}}$. In other words, in superconductors with strong electron-phonon coupling superconductivity always appears to be suppressed as temperature is lowered. The effect is particularly drastic for higher values of the exchange coupling. We therefore conclude that weak-coupling theory tends to underestimate the effect of pair breaking on superconductivity.  

\section{In-gap bound states}
Given the complicated frequency dependence of the pairing function $\Delta_n$, we would like to check whether it would be possible to have several bound states forming on impurity rather than only a single one.  Another aspect of the problem consists in elucidating how the energy of the bound state changes as one increases the strength of the electron-phonon coupling. To this end, we solve the equation for the scattering matrix, Eq. (\ref{Eq4tiwn}), numerically with the electronic propagators determined from solving the system of Eliashberg equations (\ref{ModifyEli}) by ignoring the contributions from the scattering matrix which corresponds to a case of a single impurity. 
%%%%%%%%%%%%% Fig3: Bound States - normal state %%%%%%%%%%%%%%%%%
\begin{figure}
\includegraphics[width=0.90\linewidth]{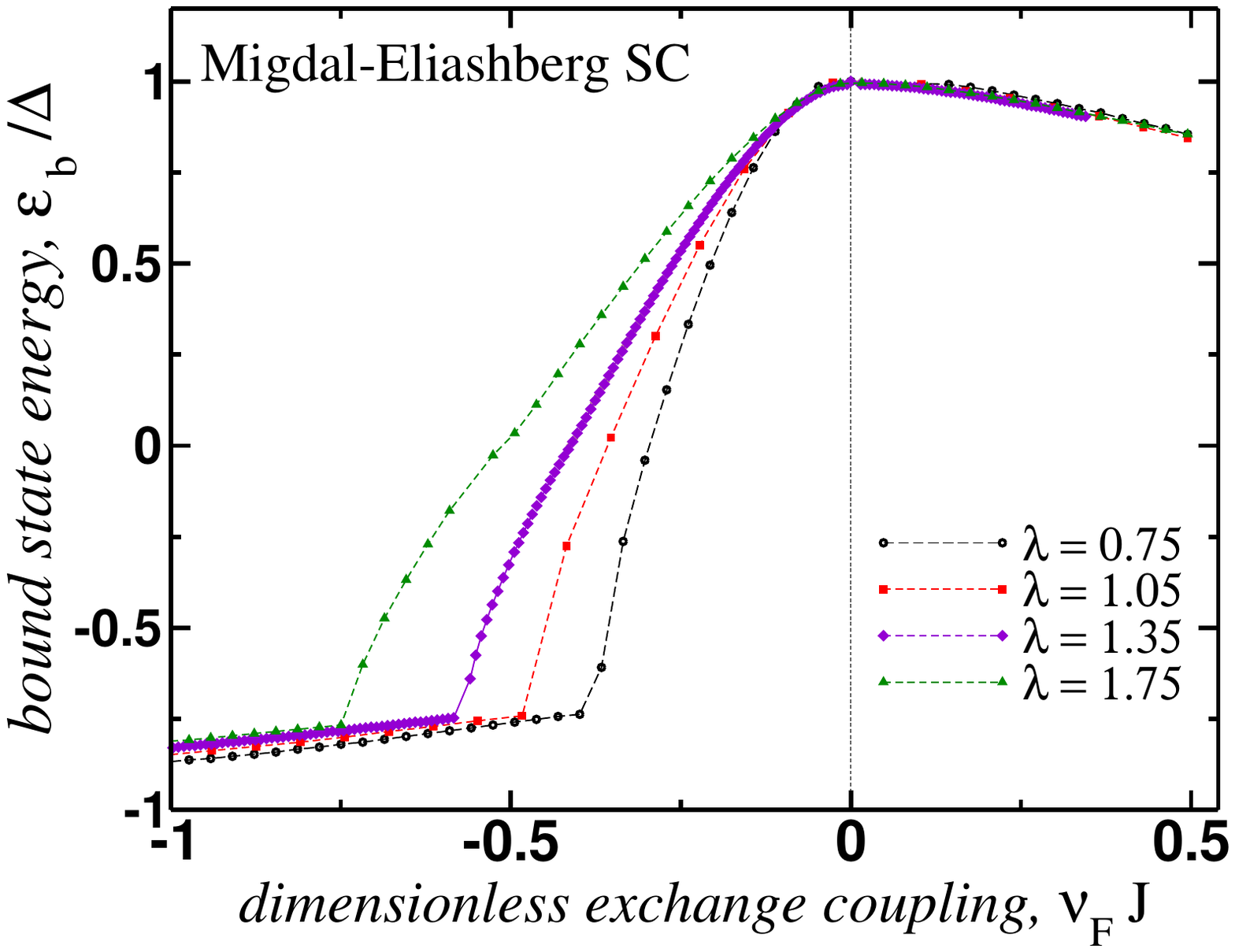}
\caption{Plot of the bound state energy as a function of dimensionless exchange coupling $\nu_FJ$ for different values of the dimensionless electron-phonon coupling $\lambda$. The dependence of the bound state energy on $\gamma$ looks qualitatively similar to the weak coupling case. The only difference arises from the fact that at larger values of $\lambda$ the bound state remains closer to the bottom of the upper Bogoliubov band. All the results shown have been obtained for temperature $T=0.01\Omega$.}
\label{Fig3-BS}
\end{figure}
%%%%%%%%%%%%%%%%%%%%%%%%%%%%%%%%%%%%%%%%%%%%%%%

We show our results in Fig. \ref{Fig3-BS}. As a reader can see, there is only one bound state, as we have expected. Furthermore, the dependence of the bound state on the exchange coupling is qualitatively similar to the one found in the weak coupling case: the bound state energy gets lower with an increase in the value of the single-impurity Kondo temperature $T_K\sim\exp(-1/\nu_F|J|)$.  The only qualitative difference that we see lies in the fact that with an increase in $\lambda$, the critical temperature of the superconducting transition also increases and, as a result, it requires higher value of the Kondo temperature for the bound state to be pushed further down from the top of the superconducting energy gap. 

\begin{figure}
\includegraphics[width=0.90\linewidth]{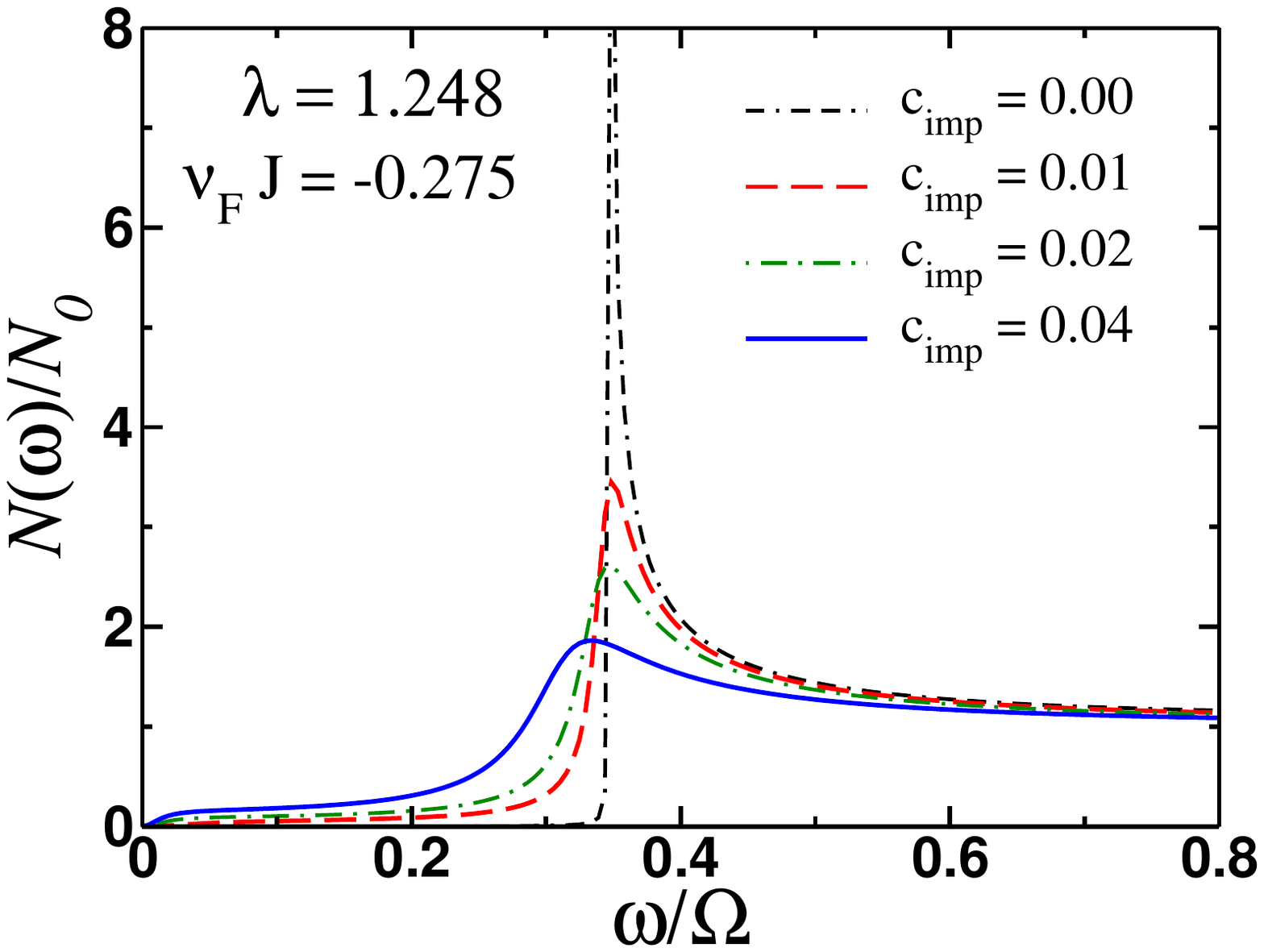}
\includegraphics[width=0.90\linewidth]{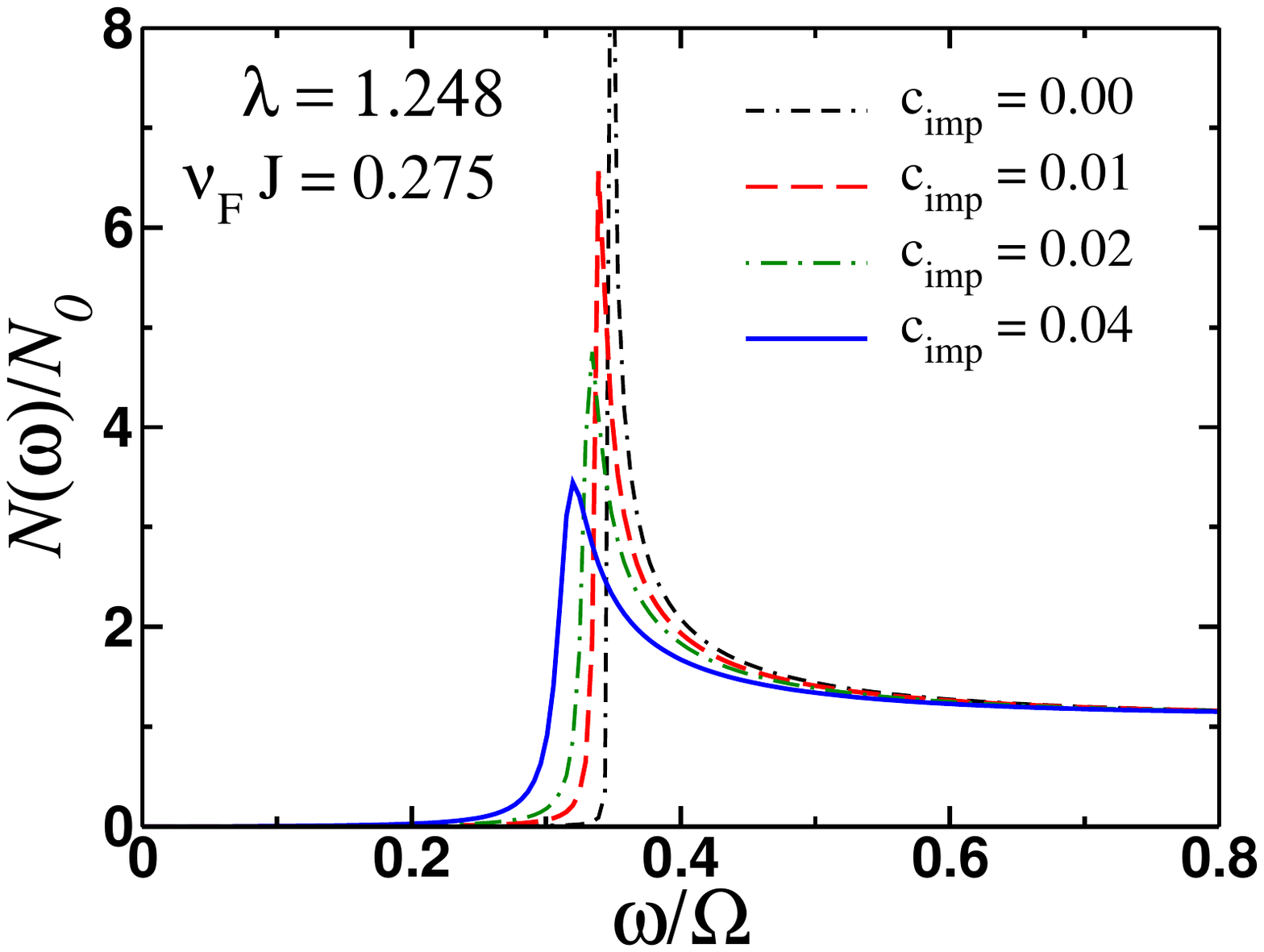}
\caption{Plot of the tunneling density of states $N(\omega)$, Eq. (\ref{Nw}), as a function of frequency for the case of antiferromagnetic exchange (top panel) and ferromagnetic exchange (bottom panel) sign of the spin exchange coupling. Both plots have been obtained for temperature $T=0.05\Omega$.}
\label{Figs45-Nw}
\end{figure}

In the context of the present discussion it is also instructive to compute the tunneling density of states which is defined by \cite{Schach1980,Schachinger1985}
\beg\label{Nw}
\frac{N(\omega)}{N_0}=\textrm{Re}\left(\frac{\omega}{\sqrt{\omega^2-\Delta^2(\omega)}}\right),
\en
where $N_0$ is the density of states of a metal in a normal state. In order to compute $N(\omega)$ we  solved equations (\ref{ModifyEli},\ref{Eq4tiwn}) self-consistently by iterations, which yields the dependence of the function 
$\Delta(i\omega_n)=\Phi(i\omega_n)/Z(i\omega_n)$ on the Matsubara frequencies $\omega_n$. Then we use the Páde approximation to compute the dependence of the pairing field on real frequency, $\Delta(i\omega_n)\to\Delta(\omega)$. The results of this calculation are presented in Fig. \ref{Figs45-Nw}.
%%%%%%%%%%%%% Fig4: Tunneling DOS %%%%%%%%%%%%%%%%%
%%%%%%%%%%%%%%%%%%%%%%%%%%%%%%%%%%%%%%%%%%%%%%%
From our results shown in Fig. \ref{Figs45-Nw} we see much stronger suppression of the superconducting 'square-root' singularity in the density of states in the case of the antiferromagnetic exchange coupling. This is in qualitative agreement with our findings for the energy of the in-gap bound states.  Indeed, the in-gap states in the case of Kondo impurities lie much deeper in the gap which leads to the broadening of the density of states into the gap. Consequently, we find that in the case of the antiferromagnetic coupling the gapless state, which corresponds to the finite value of the pairing gap and zero single-particle excitation threshold, can be achieved which less amount of magnetic impurities compared to the case of the ferromagnetic exchange coupling. 

\section{Discussion and Conclusions}
The main building block of the method we employed for our work consists in using the self-consistent equation for the scattering matrix which is obtained by writing down the equations of motion for the single particle propagator as well as vertex function. An inherent feature of such an approach is the appearance of the higher order correlation functions, which are approximated by the combination of the single particle correlation functions. Ultimately, this methodology allows one to account for the dynamical aspects of the magnetic exchange interactions. In the context of the Kondo impurity problem this procedure is approximately equivalent to a partial re-summation of the diagrams contributing to the enhancement of the antiferromagnetic exchange coupling. Therefore, one should expect it to be applicable for temperature not too far below the Kondo temperature. 
At the same time, the experimental data suggests that this method appears to be adequate even at temperatures $T\ll T_K$ as far as the calculation of the critical temperature is concerned. \cite{Maple1,Maple2,Maple3,Maple4,IanFisher2005}

Perhaps the most surprising result of our work is the dependence of the superconducting critical temperature on the impurity concentration in the case of the ferromagnetic exchange $J>0$, Eq. (\ref{Eq1}). For this case we find that superconductivity exists in a certain range of temperatures $T_{\textrm{c1}}\leq T\leq T_{\textrm{c2}}$ for a given values of impurity concentrations $c_{\textrm{imp}}$ 
with $T_{\textrm{c1}}$ vanishing in the clean limit $c_{\textrm{imp}}\to 0$. At first glance this result may seem to be the artifact of our theory. Indeed, it is well established that in the limit of weak coupling $\lambda\ll 1$ the critical temperature is gradually suppressed \cite{AG1961}. On the other hand, our results also show that starting with $J<0$ and gradually changing the value of $J$ from negative values to positive ones leads to the disappearance of superconducting state which existed at very low temperatures due to the screening of the magnetic moment thus rendering the system's ground state to become normal at $T\leq T_{\textrm{c1}}$. We thus find that while our results for the critical temperature in the case of a Kondo impurity are consistent with those found in the weak-coupling limit, $\lambda\ll 1$, our results in the case of $J>0$ are not. The origin of this discrepancy must be due to the dynamical nature of the pairing field in the case of the strong coupled superconductivity, however the full physical picture still remains obscure. 

To conclude, we have investigated how scattering on paramagnetic impurities affects strong coupling superconductivity. We have evaluated the frequency dependence of the scattering matrix, pairing field and self-energy by solving the Nagaoka and Eliashberg equations self-consistently. We then used our results to compute the energy of the in-gap bound states, critical temperature and tunneling density of states for the various values of the magnetic exchange couplings as well as dimensionless electron-phonon interaction. We find our results to be generally in agreement with those from the earlier studies on this problem. 
We hope that our findings for the dependence of the critical temperature on concentration of magnetic impurities can be verified experimentally.

\section{Acknowledgements} The authors would like to express their gratitude to Andrey Chubukov and Ilya Vekhter for useful discussions which helped to shape this study. The financial support by the National Science Foundation grant DMR-2002795 is gratefully acknowledged. Part of the work has been carried out during the Aspen Center of Physics 2024 Summer Program on "\emph{Probing Collective Excitations in Quantum Matter by Transport and Spectroscopy}", which was supported by the National Science Foundation Grant No. PHY-2210452. 

\begin{appendix}
\section{Linearized equation for the anomalous component of the scattering matrix}\label{lineqt2}
We start by introducing the following linearized quantities
\beg\label{FLin}
F_{-}(i\omega_n)=F(i\omega_n)+\delta F(i\omega_n), 
\en
where the functions appearing in the right hand side are defined according to
\beg\label{FNdF}
\begin{split}
F(i\omega_n)&=-i\nu_FJ\tilde{\omega}_n\int\limits_{-\infty}^\infty\frac{\rho(\veps)d\veps}{\veps^2+\tilde{\omega}_n^2}
\end{split}
\en
and $\delta F(i\omega_n)={\Phi(i\omega_n)}F(i\omega_n)/i\tilde{\omega}_n$.
In these formulas we introduced $\tilde{\omega}_n=\omega_nZ_n-ic_{\textrm{imp}}Jt(i\omega_n)$ and $\rho(\eps)=D^2/(\eps^2+D^2)$ ($D$ is the bandwidth).
Similarly, we write $t_{\mathrm{-}}(i\omega_n)=t(i\omega_n)+\delta t_2(i\omega_n)$, where
function $t(i\omega_n)$ is a solution of equation (\ref{Eq4tiwn}) with $\Phi_n=0$. 
Consequently, we linearize equation (\ref{Eq4tiwn}) 
and bring all the terms which contain $\delta t_2(z)$ to the left side of the equation. It obtains
\begin{widetext}
\beg\label{LinLeftSide}\nonumber
\begin{split}
&\left[\frac{S(S+1)}{4}F(z)+{\cal F}_n\left\{\frac{F(i\omega_n)-F(z)}{z-i\omega_n}t(i\omega_n)\right\}\right]\delta t(z)
+{\cal F}_n\left\{t(z)\frac{[F(i\omega_n)-F(z)]^2}{z-i\omega_n}\delta t_2(i\omega_n)\right\}\\&-
{\cal F}_n\left\{t(z)\frac{F(i\omega_n)-F(z)}{z-i\omega_n}\delta t_2(i\omega_n)\right\}=\frac{S(S+1)}{2}\left(\frac{1}{2}+F(z)t(z)\right)\delta F(z)+{\cal F}_n\left\{\frac{\delta F(i\omega_n)-\delta F(z)}{z-i\omega_n}[t(i\omega_n)-t(z)]\right\}\\&-2{\cal F}_n\left\{\frac{F(i\omega_n)-F(z)}{z-i\omega_n}[\delta F(i\omega_n)-\delta F(z)]t(i\omega_n)\right\}t(z).
\end{split}
\en
\end{widetext}
Here we took into account the equation for the function $t(i\omega_n)$. For the calculation of the critical temperature we will need to solve this equation for $\delta t(z)$ for $z=i\omega_l$. This means that equation (\ref{LinLeftSide}) reduces to the matrix equation
\beg\label{Mat4dtz}
\sum\limits_{i\omega_n}A(i\omega_l,i\omega_n)\delta t_2(i\omega_n)=\sum\limits_{i\omega_n}B(i\omega_l,i\omega_n)\Phi(i\omega_n).
\en
Thus, the task of expressing $\delta t_2(i\omega_n)$ as the linear combination of $\Phi(i\omega_n)$ reduces to inverting the matrix 
$A(i\omega_l,i\omega_n)$. The result is expression (\ref{LinExpand}) in the main text.

\end{appendix}

%\bibliography{elikondo}

\begin{thebibliography}{65}%
\makeatletter
\providecommand \@ifxundefined [1]{%
 \@ifx{#1\undefined}
}%
\providecommand \@ifnum [1]{%
 \ifnum #1\expandafter \@firstoftwo
 \else \expandafter \@secondoftwo
 \fi
}%
\providecommand \@ifx [1]{%
 \ifx #1\expandafter \@firstoftwo
 \else \expandafter \@secondoftwo
 \fi
}%
\providecommand \natexlab [1]{#1}%
\providecommand \enquote  [1]{``#1''}%
\providecommand \bibnamefont  [1]{#1}%
\providecommand \bibfnamefont [1]{#1}%
\providecommand \citenamefont [1]{#1}%
\providecommand \href@noop [0]{\@secondoftwo}%
\providecommand \href [0]{\begingroup \@sanitize@url \@href}%
\providecommand \@href[1]{\@@startlink{#1}\@@href}%
\providecommand \@@href[1]{\endgroup#1\@@endlink}%
\providecommand \@sanitize@url [0]{\catcode `\\12\catcode `\$12\catcode
  `\&12\catcode `\#12\catcode `\^12\catcode `\_12\catcode `\%12\relax}%
\providecommand \@@startlink[1]{}%
\providecommand \@@endlink[0]{}%
\providecommand \url  [0]{\begingroup\@sanitize@url \@url }%
\providecommand \@url [1]{\endgroup\@href {#1}{\urlprefix }}%
\providecommand \urlprefix  [0]{URL }%
\providecommand \Eprint [0]{\href }%
\providecommand \doibase [0]{http://dx.doi.org/}%
\providecommand \selectlanguage [0]{\@gobble}%
\providecommand \bibinfo  [0]{\@secondoftwo}%
\providecommand \bibfield  [0]{\@secondoftwo}%
\providecommand \translation [1]{[#1]}%
\providecommand \BibitemOpen [0]{}%
\providecommand \bibitemStop [0]{}%
\providecommand \bibitemNoStop [0]{.\EOS\space}%
\providecommand \EOS [0]{\spacefactor3000\relax}%
\providecommand \BibitemShut  [1]{\csname bibitem#1\endcsname}%
\let\auto@bib@innerbib\@empty
%</preamble>
\bibitem [{\citenamefont {Bucher}\ \emph {et~al.}(1975)\citenamefont {Bucher},
  \citenamefont {Maita}, \citenamefont {Hull}, \citenamefont {Fulton},\ and\
  \citenamefont {Cooper}}]{UBe13}%
  \BibitemOpen
  \bibfield  {author} {\bibinfo {author} {\bibfnamefont {E.}~\bibnamefont
  {Bucher}}, \bibinfo {author} {\bibfnamefont {J.~P.}\ \bibnamefont {Maita}},
  \bibinfo {author} {\bibfnamefont {G.~W.}\ \bibnamefont {Hull}}, \bibinfo
  {author} {\bibfnamefont {R.~C.}\ \bibnamefont {Fulton}}, \ and\ \bibinfo
  {author} {\bibfnamefont {A.~S.}\ \bibnamefont {Cooper}},\ }\bibfield  {title}
  {\enquote {\bibinfo {title} {Electronic properties of beryllides of the rare
  earth and some actinides},}\ }\href {\doibase 10.1103/PhysRevB.11.440}
  {\bibfield  {journal} {\bibinfo  {journal} {Phys. Rev. B}\ }\textbf {\bibinfo
  {volume} {11}},\ \bibinfo {pages} {440--449} (\bibinfo {year}
  {1975})}\BibitemShut {NoStop}%
\bibitem [{\citenamefont {Steglich}\ \emph {et~al.}(1979)\citenamefont
  {Steglich}, \citenamefont {Aarts}, \citenamefont {Bredl}, \citenamefont
  {Lieke}, \citenamefont {Meschede}, \citenamefont {Franz},\ and\ \citenamefont
  {Sch\"afer}}]{Ce122}%
  \BibitemOpen
  \bibfield  {author} {\bibinfo {author} {\bibfnamefont {F.}~\bibnamefont
  {Steglich}}, \bibinfo {author} {\bibfnamefont {J.}~\bibnamefont {Aarts}},
  \bibinfo {author} {\bibfnamefont {C.~D.}\ \bibnamefont {Bredl}}, \bibinfo
  {author} {\bibfnamefont {W.}~\bibnamefont {Lieke}}, \bibinfo {author}
  {\bibfnamefont {D.}~\bibnamefont {Meschede}}, \bibinfo {author}
  {\bibfnamefont {W.}~\bibnamefont {Franz}}, \ and\ \bibinfo {author}
  {\bibfnamefont {H.}~\bibnamefont {Sch\"afer}},\ }\bibfield  {title} {\enquote
  {\bibinfo {title} {Superconductivity in the presence of strong Pauli
  paramagnetism: Ce${\mathrm{Cu}}_{2}$${\mathrm{Si}}_{2}$},}\ }\href {\doibase
  10.1103/PhysRevLett.43.1892} {\bibfield  {journal} {\bibinfo  {journal}
  {Phys. Rev. Lett.}\ }\textbf {\bibinfo {volume} {43}},\ \bibinfo {pages}
  {1892--1896} (\bibinfo {year} {1979})}\BibitemShut {NoStop}%
\bibitem [{\citenamefont {Hewson}(1993)}]{Hewson1993}%
  \BibitemOpen
  \bibfield  {author} {\bibinfo {author} {\bibfnamefont {A.~C.}\ \bibnamefont
  {Hewson}},\ }\href@noop {} {\emph {\bibinfo {title} {The Kondo Problem to
  Heavy Fermions}}}\ (\bibinfo  {publisher} {Cambridge University Press,
  Cambridge, UK},\ \bibinfo {year} {1993})\BibitemShut {NoStop}%
\bibitem [{\citenamefont {Coleman}(2007)}]{Coleman2007}%
  \BibitemOpen
  \bibfield  {author} {\bibinfo {author} {\bibfnamefont {Piers}\ \bibnamefont
  {Coleman}},\ }\href {\doibase 10.1002/9780470022184.hmm105} {\emph {\bibinfo
  {title} {Handbook of Magnetism and Advanced Magnetic Materials}}}\ (\bibinfo
  {publisher} {John Wiley \& Sons, Ltd},\ \bibinfo {year} {2007})\BibitemShut
  {NoStop}%
\bibitem [{\citenamefont {Volovik}(2009)}]{Volovik2009}%
  \BibitemOpen
  \bibfield  {author} {\bibinfo {author} {\bibfnamefont {Grigory~E.}\
  \bibnamefont {Volovik}},\ }\href {\doibase
  10.1093/acprof:oso/9780199564842.001.0001} {\emph {\bibinfo {title} {{The
  Universe in a Helium Droplet}}}}\ (\bibinfo  {publisher} {Oxford University
  Press},\ \bibinfo {year} {2009})\BibitemShut {NoStop}%
\bibitem [{\citenamefont {Matthias}\ \emph {et~al.}(1963)\citenamefont
  {Matthias}, \citenamefont {Geballe},\ and\ \citenamefont
  {Compton}}]{Matthias1963}%
  \BibitemOpen
  \bibfield  {author} {\bibinfo {author} {\bibfnamefont {B.~T.}\ \bibnamefont
  {Matthias}}, \bibinfo {author} {\bibfnamefont {T.~H.}\ \bibnamefont
  {Geballe}}, \ and\ \bibinfo {author} {\bibfnamefont {V.~B.}\ \bibnamefont
  {Compton}},\ }\bibfield  {title} {\enquote {\bibinfo {title}
  {Superconductivity},}\ }\href {\doibase 10.1103/RevModPhys.35.1} {\bibfield
  {journal} {\bibinfo  {journal} {Rev. Mod. Phys.}\ }\textbf {\bibinfo {volume}
  {35}},\ \bibinfo {pages} {1--22} (\bibinfo {year} {1963})}\BibitemShut
  {NoStop}%
\bibitem [{\citenamefont {Sachdev}(2011)}]{Sachdev2011}%
  \BibitemOpen
  \bibfield  {author} {\bibinfo {author} {\bibfnamefont {Subir}\ \bibnamefont
  {Sachdev}},\ }\href@noop {} {\emph {\bibinfo {title} {Quantum Phase
  Transitions}}},\ \bibinfo {edition} {2nd}\ ed.\ (\bibinfo  {publisher}
  {Cambridge University Press},\ \bibinfo {year} {2011})\BibitemShut {NoStop}%
\bibitem [{\citenamefont {Shibauchi}\ \emph {et~al.}(2014)\citenamefont
  {Shibauchi}, \citenamefont {Carrington},\ and\ \citenamefont
  {Matsuda}}]{QCP-Review}%
  \BibitemOpen
  \bibfield  {author} {\bibinfo {author} {\bibfnamefont {T.}~\bibnamefont
  {Shibauchi}}, \bibinfo {author} {\bibfnamefont {A.}~\bibnamefont
  {Carrington}}, \ and\ \bibinfo {author} {\bibfnamefont {Y.}~\bibnamefont
  {Matsuda}},\ }\bibfield  {title} {\enquote {\bibinfo {title} {A quantum
  critical point lying beneath the superconducting dome in iron pnictides},}\
  }\href {\doibase 10.1146/annurev-conmatphys-031113-133921} {\bibfield
  {journal} {\bibinfo  {journal} {Annual Review of Condensed Matter Physics}\
  }\textbf {\bibinfo {volume} {5}},\ \bibinfo {pages} {113--135} (\bibinfo
  {year} {2014})}\BibitemShut {NoStop}%
\bibitem [{\citenamefont {Kirchner}\ \emph {et~al.}(2020)\citenamefont
  {Kirchner}, \citenamefont {Paschen}, \citenamefont {Chen}, \citenamefont
  {Wirth}, \citenamefont {Feng}, \citenamefont {Thompson},\ and\ \citenamefont
  {Si}}]{Qi-RMP20}%
  \BibitemOpen
  \bibfield  {author} {\bibinfo {author} {\bibfnamefont {Stefan}\ \bibnamefont
  {Kirchner}}, \bibinfo {author} {\bibfnamefont {Silke}\ \bibnamefont
  {Paschen}}, \bibinfo {author} {\bibfnamefont {Qiuyun}\ \bibnamefont {Chen}},
  \bibinfo {author} {\bibfnamefont {Steffen}\ \bibnamefont {Wirth}}, \bibinfo
  {author} {\bibfnamefont {Donglai}\ \bibnamefont {Feng}}, \bibinfo {author}
  {\bibfnamefont {Joe~D.}\ \bibnamefont {Thompson}}, \ and\ \bibinfo {author}
  {\bibfnamefont {Qimiao}\ \bibnamefont {Si}},\ }\bibfield  {title} {\enquote
  {\bibinfo {title} {Colloquium: Heavy-electron quantum criticality and
  single-particle spectroscopy},}\ }\href {\doibase
  10.1103/RevModPhys.92.011002} {\bibfield  {journal} {\bibinfo  {journal}
  {Rev. Mod. Phys.}\ }\textbf {\bibinfo {volume} {92}},\ \bibinfo {pages}
  {011002} (\bibinfo {year} {2020})}\BibitemShut {NoStop}%
\bibitem [{\citenamefont {Shishidou}\ \emph {et~al.}(2021)\citenamefont
  {Shishidou}, \citenamefont {Suh}, \citenamefont {Brydon}, \citenamefont
  {Weinert},\ and\ \citenamefont {Agterberg}}]{UTe2Dan}%
  \BibitemOpen
  \bibfield  {author} {\bibinfo {author} {\bibfnamefont {Tatsuya}\ \bibnamefont
  {Shishidou}}, \bibinfo {author} {\bibfnamefont {Han~Gyeol}\ \bibnamefont
  {Suh}}, \bibinfo {author} {\bibfnamefont {P.~M.~R.}\ \bibnamefont {Brydon}},
  \bibinfo {author} {\bibfnamefont {Michael}\ \bibnamefont {Weinert}}, \ and\
  \bibinfo {author} {\bibfnamefont {Daniel~F.}\ \bibnamefont {Agterberg}},\
  }\bibfield  {title} {\enquote {\bibinfo {title} {Topological band and
  superconductivity in ${\mathrm{UTe}}_{2}$},}\ }\href {\doibase
  10.1103/PhysRevB.103.104504} {\bibfield  {journal} {\bibinfo  {journal}
  {Phys. Rev. B}\ }\textbf {\bibinfo {volume} {103}},\ \bibinfo {pages}
  {104504} (\bibinfo {year} {2021})}\BibitemShut {NoStop}%
\bibitem [{\citenamefont {Aoki}\ \emph {et~al.}(2022)\citenamefont {Aoki},
  \citenamefont {Brison}, \citenamefont {Flouquet}, \citenamefont {Ishida},
  \citenamefont {Knebel}, \citenamefont {Tokunaga},\ and\ \citenamefont
  {Yanase}}]{Aoki2022}%
  \BibitemOpen
  \bibfield  {author} {\bibinfo {author} {\bibfnamefont {D}~\bibnamefont
  {Aoki}}, \bibinfo {author} {\bibfnamefont {J-P}\ \bibnamefont {Brison}},
  \bibinfo {author} {\bibfnamefont {J}~\bibnamefont {Flouquet}}, \bibinfo
  {author} {\bibfnamefont {K}~\bibnamefont {Ishida}}, \bibinfo {author}
  {\bibfnamefont {G}~\bibnamefont {Knebel}}, \bibinfo {author} {\bibfnamefont
  {Y}~\bibnamefont {Tokunaga}}, \ and\ \bibinfo {author} {\bibfnamefont
  {Y}~\bibnamefont {Yanase}},\ }\bibfield  {title} {\enquote {\bibinfo {title}
  {Unconventional superconductivity in ${\mathrm{UTe}}_{2}$},}\ }\href
  {\doibase 10.1088/1361-648x/ac5863} {\bibfield  {journal} {\bibinfo
  {journal} {Journal of Physics: Condensed Matter}\ }\textbf {\bibinfo {volume}
  {34}},\ \bibinfo {pages} {243002} (\bibinfo {year} {2022})}\BibitemShut
  {NoStop}%
\bibitem [{\citenamefont {Ran}\ \emph {et~al.}(2019)\citenamefont {Ran},
  \citenamefont {Eckberg}, \citenamefont {Ding}, \citenamefont {Furukawa},
  \citenamefont {Metz}, \citenamefont {Saha}, \citenamefont {Liu},
  \citenamefont {Zic}, \citenamefont {Kim}, \citenamefont {Paglione},\ and\
  \citenamefont {Butch}}]{UTe2Ref2}%
  \BibitemOpen
  \bibfield  {author} {\bibinfo {author} {\bibfnamefont {Sheng}\ \bibnamefont
  {Ran}}, \bibinfo {author} {\bibfnamefont {Chris}\ \bibnamefont {Eckberg}},
  \bibinfo {author} {\bibfnamefont {Qing-Ping}\ \bibnamefont {Ding}}, \bibinfo
  {author} {\bibfnamefont {Yuji}\ \bibnamefont {Furukawa}}, \bibinfo {author}
  {\bibfnamefont {Tristin}\ \bibnamefont {Metz}}, \bibinfo {author}
  {\bibfnamefont {Shanta~R.}\ \bibnamefont {Saha}}, \bibinfo {author}
  {\bibfnamefont {I-Lin}\ \bibnamefont {Liu}}, \bibinfo {author} {\bibfnamefont
  {Mark}\ \bibnamefont {Zic}}, \bibinfo {author} {\bibfnamefont {Hyunsoo}\
  \bibnamefont {Kim}}, \bibinfo {author} {\bibfnamefont {Johnpierre}\
  \bibnamefont {Paglione}}, \ and\ \bibinfo {author} {\bibfnamefont
  {Nicholas~P.}\ \bibnamefont {Butch}},\ }\bibfield  {title} {\enquote
  {\bibinfo {title} {Nearly ferromagnetic spin-triplet superconductivity},}\
  }\href {\doibase 10.1126/science.aav8645} {\bibfield  {journal} {\bibinfo
  {journal} {Science}\ }\textbf {\bibinfo {volume} {365}},\ \bibinfo {pages}
  {684--687} (\bibinfo {year} {2019})},\ \Eprint
  {http://arxiv.org/abs/https://www.science.org/doi/pdf/10.1126/science.aav8645}
  {https://www.science.org/doi/pdf/10.1126/science.aav8645} \BibitemShut
  {NoStop}%
\bibitem [{\citenamefont {de~Visser}(2019)}]{UTe2Ref3}%
  \BibitemOpen
  \bibfield  {author} {\bibinfo {author} {\bibfnamefont {Anne}\ \bibnamefont
  {de~Visser}},\ }\bibfield  {title} {\enquote {\bibinfo {title}
  {${\mathrm{UTe}}_{2}$: A new spin-triplet pairing superconductor},}\ }\href
  {\doibase 10.7566/JPSJNC.16.08} {\bibfield  {journal} {\bibinfo  {journal}
  {JPSJ News and Comments}\ }\textbf {\bibinfo {volume} {16}},\ \bibinfo
  {pages} {08} (\bibinfo {year} {2019})},\ \Eprint
  {http://arxiv.org/abs/https://doi.org/10.7566/JPSJNC.16.08}
  {https://doi.org/10.7566/JPSJNC.16.08} \BibitemShut {NoStop}%
\bibitem [{\citenamefont {Choi}\ \emph {et~al.}(2024)\citenamefont {Choi},
  \citenamefont {Lee},\ and\ \citenamefont {Yang}}]{UTe2}%
  \BibitemOpen
  \bibfield  {author} {\bibinfo {author} {\bibfnamefont {Hong~Chul}\
  \bibnamefont {Choi}}, \bibinfo {author} {\bibfnamefont {Seung~Hun}\
  \bibnamefont {Lee}}, \ and\ \bibinfo {author} {\bibfnamefont {Bohm-Jung}\
  \bibnamefont {Yang}},\ }\bibfield  {title} {\enquote {\bibinfo {title}
  {Correlated normal state fermiology and topological superconductivity in
  UTe$_2$},}\ }\href {\doibase 10.1038/s42005-024-01708-4} {\bibfield  {journal}
  {\bibinfo  {journal} {Communications Physics}\ }\textbf {\bibinfo {volume}
  {7}},\ \bibinfo {pages} {273} (\bibinfo {year} {2024})}\BibitemShut {NoStop}%
\bibitem [{\citenamefont {Mandal}\ \emph {et~al.}(2024)\citenamefont {Mandal},
  \citenamefont {Mondal}, \citenamefont {Stehno}, \citenamefont {Ili{\'c}},
  \citenamefont {Bergeret}, \citenamefont {Klapwijk}, \citenamefont {Gould},\
  and\ \citenamefont {Molenkamp}}]{FFLO2024}%
  \BibitemOpen
  \bibfield  {author} {\bibinfo {author} {\bibfnamefont {Pankaj}\ \bibnamefont
  {Mandal}}, \bibinfo {author} {\bibfnamefont {Soumi}\ \bibnamefont {Mondal}},
  \bibinfo {author} {\bibfnamefont {Martin~P.}\ \bibnamefont {Stehno}},
  \bibinfo {author} {\bibfnamefont {Stefan}\ \bibnamefont {Ili{\'c}}}, \bibinfo
  {author} {\bibfnamefont {F.~Sebastian}\ \bibnamefont {Bergeret}}, \bibinfo
  {author} {\bibfnamefont {Teun~M.}\ \bibnamefont {Klapwijk}}, \bibinfo
  {author} {\bibfnamefont {Charles}\ \bibnamefont {Gould}}, \ and\ \bibinfo
  {author} {\bibfnamefont {Laurens~W.}\ \bibnamefont {Molenkamp}},\ }\bibfield
  {title} {\enquote {\bibinfo {title} {Magnetically tunable supercurrent in
  dilute magnetic topological insulator-based Josephson junctions},}\ }\href
  {\doibase 10.1038/s41567-024-02477-1} {\bibfield  {journal} {\bibinfo
  {journal} {Nature Physics}\ }\textbf {\bibinfo {volume} {20}},\ \bibinfo
  {pages} {984--990} (\bibinfo {year} {2024})}\BibitemShut {NoStop}%
\bibitem [{\citenamefont {Putzke}\ \emph {et~al.}(2014)\citenamefont {Putzke},
  \citenamefont {Walmsley}, \citenamefont {Fletcher}, \citenamefont {Malone},
  \citenamefont {Vignolles}, \citenamefont {Proust}, \citenamefont {Badoux},
  \citenamefont {See}, \citenamefont {Beere}, \citenamefont {Ritchie},
  \citenamefont {Kasahara}, \citenamefont {Mizukami}, \citenamefont
  {Shibauchi}, \citenamefont {Matsuda},\ and\ \citenamefont
  {Carrington}}]{Putzke}%
  \BibitemOpen
  \bibfield  {author} {\bibinfo {author} {\bibfnamefont {C.}~\bibnamefont
  {Putzke}}, \bibinfo {author} {\bibfnamefont {P.}~\bibnamefont {Walmsley}},
  \bibinfo {author} {\bibfnamefont {J.~D.}\ \bibnamefont {Fletcher}}, \bibinfo
  {author} {\bibfnamefont {L.}~\bibnamefont {Malone}}, \bibinfo {author}
  {\bibfnamefont {D.}~\bibnamefont {Vignolles}}, \bibinfo {author}
  {\bibfnamefont {C.}~\bibnamefont {Proust}}, \bibinfo {author} {\bibfnamefont
  {S.}~\bibnamefont {Badoux}}, \bibinfo {author} {\bibfnamefont
  {P.}~\bibnamefont {See}}, \bibinfo {author} {\bibfnamefont {H.~E.}\
  \bibnamefont {Beere}}, \bibinfo {author} {\bibfnamefont {D.~A.}\ \bibnamefont
  {Ritchie}}, \bibinfo {author} {\bibfnamefont {S.}~\bibnamefont {Kasahara}},
  \bibinfo {author} {\bibfnamefont {Y.}~\bibnamefont {Mizukami}}, \bibinfo
  {author} {\bibfnamefont {T.}~\bibnamefont {Shibauchi}}, \bibinfo {author}
  {\bibfnamefont {Y.}~\bibnamefont {Matsuda}}, \ and\ \bibinfo {author}
  {\bibfnamefont {A.}~\bibnamefont {Carrington}},\ }\bibfield  {title}
  {\enquote {\bibinfo {title} {Anomalous critical fields in quantum critical
  superconductors},}\ }\href {\doibase 10.1038/ncomms6679} {\bibfield
  {journal} {\bibinfo  {journal} {Nature Communications}\ }\textbf {\bibinfo
  {volume} {5}},\ \bibinfo {pages} {5679} (\bibinfo {year} {2014})}\BibitemShut
  {NoStop}%
\bibitem [{\citenamefont {Joshi}\ \emph {et~al.}(2020)\citenamefont {Joshi},
  \citenamefont {Nusran}, \citenamefont {Tanatar}, \citenamefont {Cho},
  \citenamefont {Bud'ko}, \citenamefont {Canfield}, \citenamefont {Fernandes},
  \citenamefont {Levchenko},\ and\ \citenamefont {Prozorov}}]{Joshi}%
  \BibitemOpen
  \bibfield  {author} {\bibinfo {author} {\bibfnamefont {K~R}\ \bibnamefont
  {Joshi}}, \bibinfo {author} {\bibfnamefont {N~M}\ \bibnamefont {Nusran}},
  \bibinfo {author} {\bibfnamefont {M~A}\ \bibnamefont {Tanatar}}, \bibinfo
  {author} {\bibfnamefont {K}~\bibnamefont {Cho}}, \bibinfo {author}
  {\bibfnamefont {S~L}\ \bibnamefont {Bud'ko}}, \bibinfo {author}
  {\bibfnamefont {P~C}\ \bibnamefont {Canfield}}, \bibinfo {author}
  {\bibfnamefont {R~M}\ \bibnamefont {Fernandes}}, \bibinfo {author}
  {\bibfnamefont {A}~\bibnamefont {Levchenko}}, \ and\ \bibinfo {author}
  {\bibfnamefont {R}~\bibnamefont {Prozorov}},\ }\bibfield  {title} {\enquote
  {\bibinfo {title} {Quantum phase transition inside the superconducting dome
  of $\mathrm{Ba}(\mathrm{Fe}_{1-x}\mathrm{Co}_{x})_{2}\mathrm{As}_{2}$ from
  diamond-based optical magnetometry},}\ }\href {\doibase
  10.1088/1367-2630/ab85a9} {\bibfield  {journal} {\bibinfo  {journal} {New
  Journal of Physics}\ }\textbf {\bibinfo {volume} {22}},\ \bibinfo {pages}
  {053037} (\bibinfo {year} {2020})}\BibitemShut {NoStop}%
\bibitem [{\citenamefont {Lamhot}\ \emph {et~al.}(2015)\citenamefont {Lamhot},
  \citenamefont {Yagil}, \citenamefont {Shapira}, \citenamefont {Kasahara},
  \citenamefont {Watashige}, \citenamefont {Shibauchi}, \citenamefont
  {Matsuda},\ and\ \citenamefont {Auslaender}}]{Auslaender}%
  \BibitemOpen
  \bibfield  {author} {\bibinfo {author} {\bibfnamefont {Y.}~\bibnamefont
  {Lamhot}}, \bibinfo {author} {\bibfnamefont {A.}~\bibnamefont {Yagil}},
  \bibinfo {author} {\bibfnamefont {N.}~\bibnamefont {Shapira}}, \bibinfo
  {author} {\bibfnamefont {S.}~\bibnamefont {Kasahara}}, \bibinfo {author}
  {\bibfnamefont {T.}~\bibnamefont {Watashige}}, \bibinfo {author}
  {\bibfnamefont {T.}~\bibnamefont {Shibauchi}}, \bibinfo {author}
  {\bibfnamefont {Y.}~\bibnamefont {Matsuda}}, \ and\ \bibinfo {author}
  {\bibfnamefont {O.~M.}\ \bibnamefont {Auslaender}},\ }\bibfield  {title}
  {\enquote {\bibinfo {title} {Local characterization of superconductivity in
  $\mathrm{BaF}{\mathrm{e}}_{2}{(\mathrm{A}{\mathrm{s}}_{1\ensuremath{-}x}{\mathrm{P}}_{x})}_{2}$},}\
  }\href {\doibase 10.1103/PhysRevB.91.060504} {\bibfield  {journal} {\bibinfo
  {journal} {Phys. Rev. B}\ }\textbf {\bibinfo {volume} {91}},\ \bibinfo
  {pages} {060504} (\bibinfo {year} {2015})}\BibitemShut {NoStop}%
\bibitem [{\citenamefont {Hashimoto}\ \emph {et~al.}(2012)\citenamefont
  {Hashimoto}, \citenamefont {Cho}, \citenamefont {Shibauchi}, \citenamefont
  {Kasahara}, \citenamefont {Mizukami}, \citenamefont {Katsumata},
  \citenamefont {Tsuruhara}, \citenamefont {Terashima}, \citenamefont {Ikeda},
  \citenamefont {Tanatar}, \citenamefont {Kitano}, \citenamefont {Salovich},
  \citenamefont {Giannetta}, \citenamefont {Walmsley}, \citenamefont
  {Carrington}, \citenamefont {Prozorov},\ and\ \citenamefont
  {Matsuda}}]{Hashimoto1}%
  \BibitemOpen
  \bibfield  {author} {\bibinfo {author} {\bibfnamefont {K.}~\bibnamefont
  {Hashimoto}}, \bibinfo {author} {\bibfnamefont {K.}~\bibnamefont {Cho}},
  \bibinfo {author} {\bibfnamefont {T.}~\bibnamefont {Shibauchi}}, \bibinfo
  {author} {\bibfnamefont {S.}~\bibnamefont {Kasahara}}, \bibinfo {author}
  {\bibfnamefont {Y.}~\bibnamefont {Mizukami}}, \bibinfo {author}
  {\bibfnamefont {R.}~\bibnamefont {Katsumata}}, \bibinfo {author}
  {\bibfnamefont {Y.}~\bibnamefont {Tsuruhara}}, \bibinfo {author}
  {\bibfnamefont {T.}~\bibnamefont {Terashima}}, \bibinfo {author}
  {\bibfnamefont {H.}~\bibnamefont {Ikeda}}, \bibinfo {author} {\bibfnamefont
  {M.~A.}\ \bibnamefont {Tanatar}}, \bibinfo {author} {\bibfnamefont
  {H.}~\bibnamefont {Kitano}}, \bibinfo {author} {\bibfnamefont
  {N.}~\bibnamefont {Salovich}}, \bibinfo {author} {\bibfnamefont {R.~W.}\
  \bibnamefont {Giannetta}}, \bibinfo {author} {\bibfnamefont {P.}~\bibnamefont
  {Walmsley}}, \bibinfo {author} {\bibfnamefont {A.}~\bibnamefont
  {Carrington}}, \bibinfo {author} {\bibfnamefont {R.}~\bibnamefont
  {Prozorov}}, \ and\ \bibinfo {author} {\bibfnamefont {Y.}~\bibnamefont
  {Matsuda}},\ }\bibfield  {title} {\enquote {\bibinfo {title} {A sharp peak of
  the zero-temperature penetration depth at optimal composition in
  $\mathrm{BaF}{\mathrm{e}}_{2}{(\mathrm{A}{\mathrm{s}}_{1\ensuremath{-}x}{\mathrm{P}}_{x})}_{2}$},}\
  }\href {\doibase 10.1126/science.1219821} {\bibfield  {journal} {\bibinfo
  {journal} {Science}\ }\textbf {\bibinfo {volume} {336}},\ \bibinfo {pages}
  {1554--1557} (\bibinfo {year} {2012})}\BibitemShut {NoStop}%
\bibitem [{\citenamefont {Hashimoto}\ \emph {et~al.}(2013)\citenamefont
  {Hashimoto}, \citenamefont {Mizukami}, \citenamefont {Katsumata},
  \citenamefont {Shishido}, \citenamefont {Yamashita}, \citenamefont {Ikeda},
  \citenamefont {Matsuda}, \citenamefont {Schlueter}, \citenamefont {Fletcher},
  \citenamefont {Carrington}, \citenamefont {Gnida}, \citenamefont
  {Kaczorowski},\ and\ \citenamefont {Shibauchi}}]{Hashimoto2}%
  \BibitemOpen
  \bibfield  {author} {\bibinfo {author} {\bibfnamefont {Kenichiro}\
  \bibnamefont {Hashimoto}}, \bibinfo {author} {\bibfnamefont {Yuta}\
  \bibnamefont {Mizukami}}, \bibinfo {author} {\bibfnamefont {Ryo}\
  \bibnamefont {Katsumata}}, \bibinfo {author} {\bibfnamefont {Hiroaki}\
  \bibnamefont {Shishido}}, \bibinfo {author} {\bibfnamefont {Minoru}\
  \bibnamefont {Yamashita}}, \bibinfo {author} {\bibfnamefont {Hiroaki}\
  \bibnamefont {Ikeda}}, \bibinfo {author} {\bibfnamefont {Yuji}\ \bibnamefont
  {Matsuda}}, \bibinfo {author} {\bibfnamefont {John~A.}\ \bibnamefont
  {Schlueter}}, \bibinfo {author} {\bibfnamefont {Jonathan~D.}\ \bibnamefont
  {Fletcher}}, \bibinfo {author} {\bibfnamefont {Antony}\ \bibnamefont
  {Carrington}}, \bibinfo {author} {\bibfnamefont {Daniel}\ \bibnamefont
  {Gnida}}, \bibinfo {author} {\bibfnamefont {Dariusz}\ \bibnamefont
  {Kaczorowski}}, \ and\ \bibinfo {author} {\bibfnamefont {Takasada}\
  \bibnamefont {Shibauchi}},\ }\bibfield  {title} {\enquote {\bibinfo {title}
  {Anomalous superfluid density in quantum critical superconductors},}\ }\href
  {\doibase 10.1073/pnas.1221976110} {\bibfield  {journal} {\bibinfo  {journal}
  {Proceedings of the National Academy of Sciences}\ }\textbf {\bibinfo
  {volume} {110}},\ \bibinfo {pages} {3293--3297} (\bibinfo {year}
  {2013})}\BibitemShut {NoStop}%
\bibitem [{\citenamefont {Huang}\ \emph {et~al.}(2019)\citenamefont {Huang},
  \citenamefont {Li}, \citenamefont {Gao}, \citenamefont {Chen},\ and\
  \citenamefont {Zhang}}]{Huang}%
  \BibitemOpen
  \bibfield  {author} {\bibinfo {author} {\bibfnamefont {Huai-Xiang}\
  \bibnamefont {Huang}}, \bibinfo {author} {\bibfnamefont {Wei}\ \bibnamefont
  {Li}}, \bibinfo {author} {\bibfnamefont {Yi}~\bibnamefont {Gao}}, \bibinfo
  {author} {\bibfnamefont {Yan}\ \bibnamefont {Chen}}, \ and\ \bibinfo {author}
  {\bibfnamefont {Fu-Chun}\ \bibnamefont {Zhang}},\ }\bibfield  {title}
  {\enquote {\bibinfo {title} {Anomalous sharp peak in the London penetration
  depth induced by the nodeless-to-nodal superconducting transition in
  $\mathrm{BaFe}_{2}(\mathrm{As}_{1\ensuremath{-}x}\mathrm{P}_{x}{)}_{2}$},}\
  }\href {\doibase 10.1103/PhysRevB.100.144501} {\bibfield  {journal} {\bibinfo
   {journal} {Phys. Rev. B}\ }\textbf {\bibinfo {volume} {100}},\ \bibinfo
  {pages} {144501} (\bibinfo {year} {2019})}\BibitemShut {NoStop}%
\bibitem [{\citenamefont {de~Carvalho}\ \emph {et~al.}(2020)\citenamefont
  {de~Carvalho}, \citenamefont {Chubukov},\ and\ \citenamefont
  {Fernandes}}]{Carvalho}%
  \BibitemOpen
  \bibfield  {author} {\bibinfo {author} {\bibfnamefont {Vanuildo~S.}\
  \bibnamefont {de~Carvalho}}, \bibinfo {author} {\bibfnamefont {Andrey~V.}\
  \bibnamefont {Chubukov}}, \ and\ \bibinfo {author} {\bibfnamefont
  {Rafael~M.}\ \bibnamefont {Fernandes}},\ }\bibfield  {title} {\enquote
  {\bibinfo {title} {Thermodynamic signatures of an antiferromagnetic quantum
  critical point inside a superconducting dome},}\ }\href {\doibase
  10.1103/PhysRevB.102.045125} {\bibfield  {journal} {\bibinfo  {journal}
  {Phys. Rev. B}\ }\textbf {\bibinfo {volume} {102}},\ \bibinfo {pages}
  {045125} (\bibinfo {year} {2020})}\BibitemShut {NoStop}%
\bibitem [{\citenamefont {Khodas}\ \emph {et~al.}(2020)\citenamefont {Khodas},
  \citenamefont {Dzero},\ and\ \citenamefont {Levchenko}}]{Khodas}%
  \BibitemOpen
  \bibfield  {author} {\bibinfo {author} {\bibfnamefont {Maxim}\ \bibnamefont
  {Khodas}}, \bibinfo {author} {\bibfnamefont {Maxim}\ \bibnamefont {Dzero}}, \
  and\ \bibinfo {author} {\bibfnamefont {Alex}\ \bibnamefont {Levchenko}},\
  }\bibfield  {title} {\enquote {\bibinfo {title} {Anomalous thermodynamic
  properties of quantum critical superconductors},}\ }\href {\doibase
  10.1103/PhysRevB.102.184505} {\bibfield  {journal} {\bibinfo  {journal}
  {Phys. Rev. B}\ }\textbf {\bibinfo {volume} {102}},\ \bibinfo {pages}
  {184505} (\bibinfo {year} {2020})}\BibitemShut {NoStop}%
\bibitem [{\citenamefont {Levchenko}\ \emph {et~al.}(2013)\citenamefont
  {Levchenko}, \citenamefont {Vavilov}, \citenamefont {Khodas},\ and\
  \citenamefont {Chubukov}}]{Levchenko}%
  \BibitemOpen
  \bibfield  {author} {\bibinfo {author} {\bibfnamefont {A.}~\bibnamefont
  {Levchenko}}, \bibinfo {author} {\bibfnamefont {M.~G.}\ \bibnamefont
  {Vavilov}}, \bibinfo {author} {\bibfnamefont {M.}~\bibnamefont {Khodas}}, \
  and\ \bibinfo {author} {\bibfnamefont {A.~V.}\ \bibnamefont {Chubukov}},\
  }\bibfield  {title} {\enquote {\bibinfo {title} {Enhancement of the London
  penetration depth in pnictides at the onset of spin-density-wave order under
  superconducting dome},}\ }\href {\doibase 10.1103/PhysRevLett.110.177003}
  {\bibfield  {journal} {\bibinfo  {journal} {Phys. Rev. Lett.}\ }\textbf
  {\bibinfo {volume} {110}},\ \bibinfo {pages} {177003} (\bibinfo {year}
  {2013})}\BibitemShut {NoStop}%
\bibitem [{\citenamefont {Chowdhury}\ \emph {et~al.}(2013)\citenamefont
  {Chowdhury}, \citenamefont {Swingle}, \citenamefont {Berg},\ and\
  \citenamefont {Sachdev}}]{Chowdhury1}%
  \BibitemOpen
  \bibfield  {author} {\bibinfo {author} {\bibfnamefont {Debanjan}\
  \bibnamefont {Chowdhury}}, \bibinfo {author} {\bibfnamefont {Brian}\
  \bibnamefont {Swingle}}, \bibinfo {author} {\bibfnamefont {Erez}\
  \bibnamefont {Berg}}, \ and\ \bibinfo {author} {\bibfnamefont {Subir}\
  \bibnamefont {Sachdev}},\ }\bibfield  {title} {\enquote {\bibinfo {title}
  {Singularity of the London penetration depth at quantum critical points in
  superconductors},}\ }\href {\doibase 10.1103/PhysRevLett.111.157004}
  {\bibfield  {journal} {\bibinfo  {journal} {Phys. Rev. Lett.}\ }\textbf
  {\bibinfo {volume} {111}},\ \bibinfo {pages} {157004} (\bibinfo {year}
  {2013})}\BibitemShut {NoStop}%
\bibitem [{\citenamefont {Chowdhury}\ \emph {et~al.}(2015)\citenamefont
  {Chowdhury}, \citenamefont {Orenstein}, \citenamefont {Sachdev},\ and\
  \citenamefont {Senthil}}]{Chowdhury2}%
  \BibitemOpen
  \bibfield  {author} {\bibinfo {author} {\bibfnamefont {Debanjan}\
  \bibnamefont {Chowdhury}}, \bibinfo {author} {\bibfnamefont {J.}~\bibnamefont
  {Orenstein}}, \bibinfo {author} {\bibfnamefont {Subir}\ \bibnamefont
  {Sachdev}}, \ and\ \bibinfo {author} {\bibfnamefont {T.}~\bibnamefont
  {Senthil}},\ }\bibfield  {title} {\enquote {\bibinfo {title} {Phase
  transition beneath the superconducting dome in
  $\mathrm{BaFe}_{2}{({\text{As}}_{1\ensuremath{-}x}{\text{P}}_{x})}_{2}$},}\
  }\href {\doibase 10.1103/PhysRevB.92.081113} {\bibfield  {journal} {\bibinfo
  {journal} {Phys. Rev. B}\ }\textbf {\bibinfo {volume} {92}},\ \bibinfo
  {pages} {081113} (\bibinfo {year} {2015})}\BibitemShut {NoStop}%
\bibitem [{\citenamefont {Jujo}(2018)}]{Jujo2018}%
  \BibitemOpen
  \bibfield  {author} {\bibinfo {author} {\bibfnamefont {Takanobu}\
  \bibnamefont {Jujo}},\ }\bibfield  {title} {\enquote {\bibinfo {title}
  {Two-photon absorption and amplitude mode in conventional superconductors
  with paramagnetic impurities},}\ }\href {\doibase
  10.1088/1742-6596/969/1/012035} {\bibfield  {journal} {\bibinfo  {journal}
  {Journal of Physics: Conference Series}\ }\textbf {\bibinfo {volume} {969}},\
  \bibinfo {pages} {012035} (\bibinfo {year} {2018})}\BibitemShut {NoStop}%
\bibitem [{\citenamefont {Li}\ and\ \citenamefont {Dzero}(2024)}]{Higgs2024}%
  \BibitemOpen
  \bibfield  {author} {\bibinfo {author} {\bibfnamefont {Yantao}\ \bibnamefont
  {Li}}\ and\ \bibinfo {author} {\bibfnamefont {Maxim}\ \bibnamefont {Dzero}},\
  }\bibfield  {title} {\enquote {\bibinfo {title} {Amplitude Higgs mode in
  superconductors with magnetic impurities},}\ }\href {\doibase
  10.1103/PhysRevB.109.054520} {\bibfield  {journal} {\bibinfo  {journal}
  {Phys. Rev. B}\ }\textbf {\bibinfo {volume} {109}},\ \bibinfo {pages}
  {054520} (\bibinfo {year} {2024})}\BibitemShut {NoStop}%
\bibitem [{\citenamefont {Yang}\ and\ \citenamefont
  {Wu}(2024)}]{LinResponse2024}%
  \BibitemOpen
  \bibfield  {author} {\bibinfo {author} {\bibfnamefont {F.}~\bibnamefont
  {Yang}}\ and\ \bibinfo {author} {\bibfnamefont {M.~W.}\ \bibnamefont {Wu}},\
  }\bibfield  {title} {\enquote {\bibinfo {title} {Diamagnetic property and
  optical absorption of conventional superconductors with magnetic impurities
  in linear response},}\ }\href {\doibase 10.1103/PhysRevB.109.064508}
  {\bibfield  {journal} {\bibinfo  {journal} {Phys. Rev. B}\ }\textbf {\bibinfo
  {volume} {109}},\ \bibinfo {pages} {064508} (\bibinfo {year}
  {2024})}\BibitemShut {NoStop}%
\bibitem [{\citenamefont {Chubukov}(2012)}]{ChubukovAnRev2012}%
  \BibitemOpen
  \bibfield  {author} {\bibinfo {author} {\bibfnamefont {Andrey}\ \bibnamefont
  {Chubukov}},\ }\bibfield  {title} {\enquote {\bibinfo {title} {Pairing
  mechanism in Fe-based superconductors},}\ }\href {\doibase
  https://doi.org/10.1146/annurev-conmatphys-020911-125055} {\bibfield
  {journal} {\bibinfo  {journal} {Annual Review of Condensed Matter Physics}\
  }\textbf {\bibinfo {volume} {3}},\ \bibinfo {pages} {57--92} (\bibinfo {year}
  {2012})}\BibitemShut {NoStop}%
\bibitem [{\citenamefont {Si}\ and\ \citenamefont {Hussey}(2023)}]{QSi2023}%
  \BibitemOpen
  \bibfield  {author} {\bibinfo {author} {\bibfnamefont {Qimiao}\ \bibnamefont
  {Si}}\ and\ \bibinfo {author} {\bibfnamefont {Nigel~E.}\ \bibnamefont
  {Hussey}},\ }\bibfield  {title} {\enquote {\bibinfo {title} {{Iron-based
  superconductors: Teenage, complex, challenging}},}\ }\href {\doibase
  10.1063/PT.3.5235} {\bibfield  {journal} {\bibinfo  {journal} {Physics
  Today}\ }\textbf {\bibinfo {volume} {76}},\ \bibinfo {pages} {34--40}
  (\bibinfo {year} {2023})},\ \Eprint
  {http://arxiv.org/abs/https://pubs.aip.org/physicstoday/article-pdf/76/5/34/20086106/34\_1\_pt.3.5235.pdf}
  {https://pubs.aip.org/physicstoday/article-pdf/76/5/34/20086106/34\_1\_pt.3.5235.pdf}
  \BibitemShut {NoStop}%
\bibitem [{\citenamefont {Dzero}\ \emph {et~al.}(2015)\citenamefont {Dzero},
  \citenamefont {Khodas}, \citenamefont {Klironomos}, \citenamefont {Vavilov},\
  and\ \citenamefont {Levchenko}}]{Dzero}%
  \BibitemOpen
  \bibfield  {author} {\bibinfo {author} {\bibfnamefont {M.}~\bibnamefont
  {Dzero}}, \bibinfo {author} {\bibfnamefont {M.}~\bibnamefont {Khodas}},
  \bibinfo {author} {\bibfnamefont {A.~D.}\ \bibnamefont {Klironomos}},
  \bibinfo {author} {\bibfnamefont {M.~G.}\ \bibnamefont {Vavilov}}, \ and\
  \bibinfo {author} {\bibfnamefont {A.}~\bibnamefont {Levchenko}},\ }\bibfield
  {title} {\enquote {\bibinfo {title} {Magnetic penetration depth in disordered
  iron-based superconductors},}\ }\href {\doibase 10.1103/PhysRevB.92.144501}
  {\bibfield  {journal} {\bibinfo  {journal} {Phys. Rev. B}\ }\textbf {\bibinfo
  {volume} {92}},\ \bibinfo {pages} {144501} (\bibinfo {year}
  {2015})}\BibitemShut {NoStop}%
\bibitem [{\citenamefont {Abrikosov}\ and\ \citenamefont
  {Gor'kov}(1961)}]{AG1961}%
  \BibitemOpen
  \bibfield  {author} {\bibinfo {author} {\bibfnamefont {A.~A.}\ \bibnamefont
  {Abrikosov}}\ and\ \bibinfo {author} {\bibfnamefont {L.~P.}\ \bibnamefont
  {Gor'kov}},\ }\bibfield  {title} {\enquote {\bibinfo {title} {Contribution to
  the theory of superconducting alloys with paramagnetic impurities},}\
  }\href@noop {} {\bibfield  {journal} {\bibinfo  {journal} {Sov. Phys. -
  JETP}\ }\textbf {\bibinfo {volume} {12}},\ \bibinfo {pages} {1243} (\bibinfo
  {year} {1961})}\BibitemShut {NoStop}%
\bibitem [{\citenamefont {Bardeen}\ \emph {et~al.}(1957)\citenamefont
  {Bardeen}, \citenamefont {Cooper},\ and\ \citenamefont
  {Schrieffer}}]{bcstheory}%
  \BibitemOpen
  \bibfield  {author} {\bibinfo {author} {\bibfnamefont {J}~\bibnamefont
  {Bardeen}}, \bibinfo {author} {\bibfnamefont {L~N}\ \bibnamefont {Cooper}}, \
  and\ \bibinfo {author} {\bibfnamefont {J~R}\ \bibnamefont {Schrieffer}},\
  }\bibfield  {title} {\enquote {\bibinfo {title} {{\sl Theory of
  Superconductivity}},}\ }\href@noop {} {\bibfield  {journal} {\bibinfo
  {journal} {Physical Review}\ }\textbf {\bibinfo {volume} {108}},\ \bibinfo
  {pages} {1175--1204} (\bibinfo {year} {1957})}\BibitemShut {NoStop}%
\bibitem [{\citenamefont {Yu}(1965)}]{Yu}%
  \BibitemOpen
  \bibfield  {author} {\bibinfo {author} {\bibfnamefont {L.}~\bibnamefont
  {Yu}},\ }\bibfield  {title} {\enquote {\bibinfo {title} {Classical spins in
  superconductors},}\ }\href@noop {} {\bibfield  {journal} {\bibinfo  {journal}
  {Acta Phys. Sin.}\ }\textbf {\bibinfo {volume} {21}},\ \bibinfo {pages} {75}
  (\bibinfo {year} {1965})}\BibitemShut {NoStop}%
\bibitem [{\citenamefont {Shiba}(1968)}]{Shiba}%
  \BibitemOpen
  \bibfield  {author} {\bibinfo {author} {\bibfnamefont {H.}~\bibnamefont
  {Shiba}},\ }\bibfield  {title} {\enquote {\bibinfo {title} {Classical spins
  in superconductors},}\ }\href@noop {} {\bibfield  {journal} {\bibinfo
  {journal} {Prog. Theor. Phys.}\ }\textbf {\bibinfo {volume} {40}},\ \bibinfo
  {pages} {435} (\bibinfo {year} {1968})}\BibitemShut {NoStop}%
\bibitem [{\citenamefont {Rusinov}(1969)}]{Rusinov}%
  \BibitemOpen
  \bibfield  {author} {\bibinfo {author} {\bibfnamefont {A.~I.}\ \bibnamefont
  {Rusinov}},\ }\bibfield  {title} {\enquote {\bibinfo {title} {Theory of
  gapless superconductivity in alloys containing paramagnetic impurities},}\
  }\href@noop {} {\bibfield  {journal} {\bibinfo  {journal} {Sov. Phys. JETP}\
  }\textbf {\bibinfo {volume} {429}},\ \bibinfo {pages} {1101} (\bibinfo {year}
  {1969})}\BibitemShut {NoStop}%
\bibitem [{\citenamefont {Balatsky}\ \emph {et~al.}(2006)\citenamefont
  {Balatsky}, \citenamefont {Vekhter},\ and\ \citenamefont
  {Zhu}}]{Balatsky-RMP}%
  \BibitemOpen
  \bibfield  {author} {\bibinfo {author} {\bibfnamefont {A.~V.}\ \bibnamefont
  {Balatsky}}, \bibinfo {author} {\bibfnamefont {I.}~\bibnamefont {Vekhter}}, \
  and\ \bibinfo {author} {\bibfnamefont {Jian-Xin}\ \bibnamefont {Zhu}},\
  }\bibfield  {title} {\enquote {\bibinfo {title} {Impurity-induced states in
  conventional and unconventional superconductors},}\ }\href {\doibase
  10.1103/RevModPhys.78.373} {\bibfield  {journal} {\bibinfo  {journal} {Rev.
  Mod. Phys.}\ }\textbf {\bibinfo {volume} {78}},\ \bibinfo {pages} {373--433}
  (\bibinfo {year} {2006})}\BibitemShut {NoStop}%
\bibitem [{\citenamefont {Galitski}\ and\ \citenamefont
  {Larkin}(2002)}]{VityaG-2002}%
  \BibitemOpen
  \bibfield  {author} {\bibinfo {author} {\bibfnamefont {V.~M.}\ \bibnamefont
  {Galitski}}\ and\ \bibinfo {author} {\bibfnamefont {A.~I.}\ \bibnamefont
  {Larkin}},\ }\bibfield  {title} {\enquote {\bibinfo {title} {Spin glass
  versus superconductivity},}\ }\href {\doibase 10.1103/PhysRevB.66.064526}
  {\bibfield  {journal} {\bibinfo  {journal} {Phys. Rev. B}\ }\textbf {\bibinfo
  {volume} {66}},\ \bibinfo {pages} {064526} (\bibinfo {year}
  {2002})}\BibitemShut {NoStop}%
\bibitem [{\citenamefont {Kondo}(1964)}]{Kondo64}%
  \BibitemOpen
  \bibfield  {author} {\bibinfo {author} {\bibfnamefont {J.}~\bibnamefont
  {Kondo}},\ }\bibfield  {title} {\enquote {\bibinfo {title} {{\sl Resistance
  Minimum in Dilute Magnetic Alloys}},}\ }\href@noop {} {\bibfield  {journal}
  {\bibinfo  {journal} {Prog. Theor. Phys.}\ }\textbf {\bibinfo {volume}
  {32}},\ \bibinfo {pages} {37--49} (\bibinfo {year} {1964})}\BibitemShut
  {NoStop}%
\bibitem [{\citenamefont {Nagaoka}(1965)}]{Nagaoka1965}%
  \BibitemOpen
  \bibfield  {author} {\bibinfo {author} {\bibfnamefont {Yosuke}\ \bibnamefont
  {Nagaoka}},\ }\bibfield  {title} {\enquote {\bibinfo {title} {Self-consistent
  treatment of Kondo effect in dilute alloys},}\ }\href {\doibase
  10.1103/PhysRev.138.A1112} {\bibfield  {journal} {\bibinfo  {journal} {Phys.
  Rev.}\ }\textbf {\bibinfo {volume} {138}},\ \bibinfo {pages} {A1112--A1120}
  (\bibinfo {year} {1965})}\BibitemShut {NoStop}%
\bibitem [{\citenamefont {Zittartz}\ and\ \citenamefont
  {M{\"u}ller-Hartmann}(1968)}]{MHZNormal}%
  \BibitemOpen
  \bibfield  {author} {\bibinfo {author} {\bibfnamefont {J.}~\bibnamefont
  {Zittartz}}\ and\ \bibinfo {author} {\bibfnamefont {E.}~\bibnamefont
  {M{\"u}ller-Hartmann}},\ }\bibfield  {title} {\enquote {\bibinfo {title}
  {Green's function theory of the Kondo effect in dilute magnetic alloys},}\
  }\href {\doibase 10.1007/BF01380112} {\bibfield  {journal} {\bibinfo
  {journal} {Zeitschrift f{\"u}r Physik}\ }\textbf {\bibinfo {volume} {212}},\
  \bibinfo {pages} {380--407} (\bibinfo {year} {1968})}\BibitemShut {NoStop}%
\bibitem [{\citenamefont {Abrikosov}(1988)}]{Abrikosov1988}%
  \BibitemOpen
  \bibfield  {author} {\bibinfo {author} {\bibfnamefont {A.~A.}\ \bibnamefont
  {Abrikosov}},\ }\href@noop {} {\emph {\bibinfo {title} {{Fundamentals of the
  theory of metals}}}}\ (\bibinfo  {publisher} {North-Holland, Amsterdam},\
  \bibinfo {year} {1988})\BibitemShut {NoStop}%
\bibitem [{\citenamefont {Zittartz}\ and\ \citenamefont
  {M{\"u}ller-Hartmann}(1970)}]{MHZ1970I}%
  \BibitemOpen
  \bibfield  {author} {\bibinfo {author} {\bibfnamefont {J.}~\bibnamefont
  {Zittartz}}\ and\ \bibinfo {author} {\bibfnamefont {E.}~\bibnamefont
  {M{\"u}ller-Hartmann}},\ }\bibfield  {title} {\enquote {\bibinfo {title}
  {Theory of magnetic impurities in superconductors \textrm{I}},}\ }\href
  {\doibase 10.1007/BF01394943} {\bibfield  {journal} {\bibinfo  {journal}
  {Zeitschrift f{\"u}r Physik}\ }\textbf {\bibinfo {volume} {232}},\ \bibinfo
  {pages} {11--31} (\bibinfo {year} {1970})}\BibitemShut {NoStop}%
\bibitem [{\citenamefont {M\"uller-Hartmann}\ and\ \citenamefont
  {Zittartz}(1971)}]{MHZ1971}%
  \BibitemOpen
  \bibfield  {author} {\bibinfo {author} {\bibfnamefont {E.}~\bibnamefont
  {M\"uller-Hartmann}}\ and\ \bibinfo {author} {\bibfnamefont {J.}~\bibnamefont
  {Zittartz}},\ }\bibfield  {title} {\enquote {\bibinfo {title} {Kondo effect
  in superconductors},}\ }\href {\doibase 10.1103/PhysRevLett.26.428}
  {\bibfield  {journal} {\bibinfo  {journal} {Phys. Rev. Lett.}\ }\textbf
  {\bibinfo {volume} {26}},\ \bibinfo {pages} {428--432} (\bibinfo {year}
  {1971})}\BibitemShut {NoStop}%
\bibitem [{\citenamefont {M{\"u}ller-Hartmann}\ and\ \citenamefont
  {Zittartz}(1970)}]{MHZ1970II}%
  \BibitemOpen
  \bibfield  {author} {\bibinfo {author} {\bibfnamefont {E.}~\bibnamefont
  {M{\"u}ller-Hartmann}}\ and\ \bibinfo {author} {\bibfnamefont
  {J.}~\bibnamefont {Zittartz}},\ }\bibfield  {title} {\enquote {\bibinfo
  {title} {Theory of magnetic impurities in superconductors \textrm{II}},}\
  }\href {\doibase 10.1007/BF01392497} {\bibfield  {journal} {\bibinfo
  {journal} {Zeitschrift f{\"u}r Physik}\ }\textbf {\bibinfo {volume} {234}},\
  \bibinfo {pages} {58--69} (\bibinfo {year} {1970})}\BibitemShut {NoStop}%
\bibitem [{\citenamefont {Schuh}\ and\ \citenamefont
  {M{\"u}ller-Hartmann}(1978)}]{Schuh1978}%
  \BibitemOpen
  \bibfield  {author} {\bibinfo {author} {\bibfnamefont {B.}~\bibnamefont
  {Schuh}}\ and\ \bibinfo {author} {\bibfnamefont {E.}~\bibnamefont
  {M{\"u}ller-Hartmann}},\ }\bibfield  {title} {\enquote {\bibinfo {title}
  {Self-consistent theory of pair-breaking in Kondo superconductors},}\ }\href
  {\doibase 10.1007/BF01354836} {\bibfield  {journal} {\bibinfo  {journal}
  {Zeitschrift f{\"u}r Physik B Condensed Matter}\ }\textbf {\bibinfo {volume}
  {29}},\ \bibinfo {pages} {39--46} (\bibinfo {year} {1978})}\BibitemShut
  {NoStop}%
\bibitem [{\citenamefont {Matsuura}\ \emph {et~al.}(1977)\citenamefont
  {Matsuura}, \citenamefont {Ichinose},\ and\ \citenamefont {Nagaoka}}]{MIN}%
  \BibitemOpen
  \bibfield  {author} {\bibinfo {author} {\bibfnamefont {Tamifusa}\
  \bibnamefont {Matsuura}}, \bibinfo {author} {\bibfnamefont {Shin'ichi}\
  \bibnamefont {Ichinose}}, \ and\ \bibinfo {author} {\bibfnamefont {Yosuke}\
  \bibnamefont {Nagaoka}},\ }\bibfield  {title} {\enquote {\bibinfo {title}
  {{Theory of Kondo Effect in Superconductors. I: Transition Temperature and
  Upper Critical Field}},}\ }\href {\doibase 10.1143/PTP.57.713} {\bibfield
  {journal} {\bibinfo  {journal} {Progress of Theoretical Physics}\ }\textbf
  {\bibinfo {volume} {57}},\ \bibinfo {pages} {713--733} (\bibinfo {year}
  {1977})},\ \Eprint
  {http://arxiv.org/abs/https://academic.oup.com/ptp/article-pdf/57/3/713/5466656/57-3-713.pdf}
  {https://academic.oup.com/ptp/article-pdf/57/3/713/5466656/57-3-713.pdf}
  \BibitemShut {NoStop}%
\bibitem [{\citenamefont {Maple}\ \emph {et~al.}(1972)\citenamefont {Maple},
  \citenamefont {Fertig}, \citenamefont {Mota}, \citenamefont {DeLong},
  \citenamefont {Wohlleben},\ and\ \citenamefont {Fitzgerald}}]{Maple1}%
  \BibitemOpen
  \bibfield  {author} {\bibinfo {author} {\bibfnamefont {M.B.}\ \bibnamefont
  {Maple}}, \bibinfo {author} {\bibfnamefont {W.A.}\ \bibnamefont {Fertig}},
  \bibinfo {author} {\bibfnamefont {A.C.}\ \bibnamefont {Mota}}, \bibinfo
  {author} {\bibfnamefont {L.E.}\ \bibnamefont {DeLong}}, \bibinfo {author}
  {\bibfnamefont {D.}~\bibnamefont {Wohlleben}}, \ and\ \bibinfo {author}
  {\bibfnamefont {R.}~\bibnamefont {Fitzgerald}},\ }\bibfield  {title}
  {\enquote {\bibinfo {title} {The re-entrant superconducting-normal phase
  boundary of the Kondo system (La, Ce)Al$_2$},}\ }\href {\doibase
  https://doi.org/10.1016/0038-1098(72)90281-5} {\bibfield  {journal} {\bibinfo
   {journal} {Solid State Communications}\ }\textbf {\bibinfo {volume} {11}},\
  \bibinfo {pages} {829--834} (\bibinfo {year} {1972})}\BibitemShut {NoStop}%
\bibitem [{\citenamefont {Huber}\ \emph {et~al.}(1974)\citenamefont {Huber},
  \citenamefont {Fertig},\ and\ \citenamefont {Maple}}]{Maple2}%
  \BibitemOpen
  \bibfield  {author} {\bibinfo {author} {\bibfnamefont {J.G.}\ \bibnamefont
  {Huber}}, \bibinfo {author} {\bibfnamefont {W.A.}\ \bibnamefont {Fertig}}, \
  and\ \bibinfo {author} {\bibfnamefont {M.B.}\ \bibnamefont {Maple}},\
  }\bibfield  {title} {\enquote {\bibinfo {title} {Superconducting—normal
  phase boundaries of (La, Th)Ce systems},}\ }\href {\doibase
  https://doi.org/10.1016/0038-1098(74)91119-3} {\bibfield  {journal} {\bibinfo
   {journal} {Solid State Communications}\ }\textbf {\bibinfo {volume} {15}},\
  \bibinfo {pages} {453--457} (\bibinfo {year} {1974})}\BibitemShut {NoStop}%
\bibitem [{\citenamefont {Winzer}(1977)}]{Maple3}%
  \BibitemOpen
  \bibfield  {author} {\bibinfo {author} {\bibfnamefont {K.}~\bibnamefont
  {Winzer}},\ }\bibfield  {title} {\enquote {\bibinfo {title} {Evidence for a
  three $T_c$ behavior of the Kondo superconductor (La, Y)Ce},}\ }\href {\doibase
  https://doi.org/10.1016/0038-1098(77)90161-2} {\bibfield  {journal} {\bibinfo
   {journal} {Solid State Communications}\ }\textbf {\bibinfo {volume} {24}},\
  \bibinfo {pages} {551--556} (\bibinfo {year} {1977})}\BibitemShut {NoStop}%
\bibitem [{\citenamefont {Dreyer}\ \emph {et~al.}(1982)\citenamefont {Dreyer},
  \citenamefont {Krug},\ and\ \citenamefont {Winzer}}]{Maple4}%
  \BibitemOpen
  \bibfield  {author} {\bibinfo {author} {\bibfnamefont {R.}~\bibnamefont
  {Dreyer}}, \bibinfo {author} {\bibfnamefont {T.}~\bibnamefont {Krug}}, \ and\
  \bibinfo {author} {\bibfnamefont {K.}~\bibnamefont {Winzer}},\ }\bibfield
  {title} {\enquote {\bibinfo {title} {Two-$T_c$ and three-$T_c$ behavior of the Kondo
  superconductor (La, Y)Ce},}\ }\href {\doibase 10.1007/BF00681721} {\bibfield
  {journal} {\bibinfo  {journal} {Journal of Low Temperature Physics}\ }\textbf
  {\bibinfo {volume} {48}},\ \bibinfo {pages} {111--124} (\bibinfo {year}
  {1982})}\BibitemShut {NoStop}%
\bibitem [{\citenamefont {Dzero}\ and\ \citenamefont
  {Schmalian}(2005)}]{DzeroCK}%
  \BibitemOpen
  \bibfield  {author} {\bibinfo {author} {\bibfnamefont {Maxim}\ \bibnamefont
  {Dzero}}\ and\ \bibinfo {author} {\bibfnamefont {J\"org}\ \bibnamefont
  {Schmalian}},\ }\bibfield  {title} {\enquote {\bibinfo {title}
  {Superconductivity in charge Kondo systems},}\ }\href {\doibase
  10.1103/PhysRevLett.94.157003} {\bibfield  {journal} {\bibinfo  {journal}
  {Phys. Rev. Lett.}\ }\textbf {\bibinfo {volume} {94}},\ \bibinfo {pages}
  {157003} (\bibinfo {year} {2005})}\BibitemShut {NoStop}%
\bibitem [{\citenamefont {Matsushita}\ \emph {et~al.}(2005)\citenamefont
  {Matsushita}, \citenamefont {Bluhm}, \citenamefont {Geballe},\ and\
  \citenamefont {Fisher}}]{IanFisher2005}%
  \BibitemOpen
  \bibfield  {author} {\bibinfo {author} {\bibfnamefont {Y.}~\bibnamefont
  {Matsushita}}, \bibinfo {author} {\bibfnamefont {H.}~\bibnamefont {Bluhm}},
  \bibinfo {author} {\bibfnamefont {T.~H.}\ \bibnamefont {Geballe}}, \ and\
  \bibinfo {author} {\bibfnamefont {I.~R.}\ \bibnamefont {Fisher}},\ }\bibfield
   {title} {\enquote {\bibinfo {title} {Evidence for charge Kondo effect in
  superconducting tl-doped pbte},}\ }\href {\doibase
  10.1103/PhysRevLett.94.157002} {\bibfield  {journal} {\bibinfo  {journal}
  {Phys. Rev. Lett.}\ }\textbf {\bibinfo {volume} {94}},\ \bibinfo {pages}
  {157002} (\bibinfo {year} {2005})}\BibitemShut {NoStop}%
\bibitem [{\citenamefont {Barzykin}\ and\ \citenamefont
  {Gor'kov}(2005)}]{LPG2005}%
  \BibitemOpen
  \bibfield  {author} {\bibinfo {author} {\bibfnamefont {Victor}\ \bibnamefont
  {Barzykin}}\ and\ \bibinfo {author} {\bibfnamefont {L.~P.}\ \bibnamefont
  {Gor'kov}},\ }\bibfield  {title} {\enquote {\bibinfo {title} {Competition
  between phonon superconductivity and Kondo screening in mixed valence and
  heavy fermion compounds},}\ }\href {\doibase 10.1103/PhysRevB.71.214521}
  {\bibfield  {journal} {\bibinfo  {journal} {Phys. Rev. B}\ }\textbf {\bibinfo
  {volume} {71}},\ \bibinfo {pages} {214521} (\bibinfo {year}
  {2005})}\BibitemShut {NoStop}%
\bibitem [{\citenamefont {Awelewa}\ and\ \citenamefont
  {Dzero}(2024)}]{Awelewa2024}%
  \BibitemOpen
  \bibfield  {author} {\bibinfo {author} {\bibfnamefont {Samuel}\ \bibnamefont
  {Awelewa}}\ and\ \bibinfo {author} {\bibfnamefont {Maxim}\ \bibnamefont
  {Dzero}},\ }\bibfield  {title} {\enquote {\bibinfo {title}
  {Migdal–Eliashberg superconductivity in a Kondo lattice},}\ }\href
  {\doibase 10.1088/1361-648X/ad43a5} {\bibfield  {journal} {\bibinfo
  {journal} {Journal of Physics: Condensed Matter}\ }\textbf {\bibinfo {volume}
  {36}},\ \bibinfo {pages} {325602} (\bibinfo {year} {2024})}\BibitemShut
  {NoStop}%
\bibitem [{\citenamefont {Schachinger}\ \emph {et~al.}(1980)\citenamefont
  {Schachinger}, \citenamefont {Daams},\ and\ \citenamefont
  {Carbotte}}]{Schach1980}%
  \BibitemOpen
  \bibfield  {author} {\bibinfo {author} {\bibfnamefont {E.}~\bibnamefont
  {Schachinger}}, \bibinfo {author} {\bibfnamefont {J.~M.}\ \bibnamefont
  {Daams}}, \ and\ \bibinfo {author} {\bibfnamefont {J.~P.}\ \bibnamefont
  {Carbotte}},\ }\bibfield  {title} {\enquote {\bibinfo {title} {Paramagnetic
  impurities in strong coupling superconductors},}\ }\href {\doibase
  10.1103/PhysRevB.22.3194} {\bibfield  {journal} {\bibinfo  {journal} {Phys.
  Rev. B}\ }\textbf {\bibinfo {volume} {22}},\ \bibinfo {pages} {3194--3199}
  (\bibinfo {year} {1980})}\BibitemShut {NoStop}%
\bibitem [{\citenamefont {Schachinger}\ and\ \citenamefont
  {Wunder}(1985)}]{Schachinger1985}%
  \BibitemOpen
  \bibfield  {author} {\bibinfo {author} {\bibfnamefont {E.}~\bibnamefont
  {Schachinger}}\ and\ \bibinfo {author} {\bibfnamefont {R.}~\bibnamefont
  {Wunder}},\ }\bibfield  {title} {\enquote {\bibinfo {title} {Strong coupling
  Kondo superconductors},}\ }\href {\doibase
  https://doi.org/10.1016/0378-4363(85)90532-7} {\bibfield  {journal} {\bibinfo
   {journal} {Physica B+C}\ }\textbf {\bibinfo {volume} {135}},\ \bibinfo
  {pages} {464--467} (\bibinfo {year} {1985})}\BibitemShut {NoStop}%
\bibitem [{\citenamefont {Schachinger}(1982)}]{Schach1982}%
  \BibitemOpen
  \bibfield  {author} {\bibinfo {author} {\bibfnamefont {E.}~\bibnamefont
  {Schachinger}},\ }\bibfield  {title} {\enquote {\bibinfo {title} {Strong
  coupling theory of superconducting alloys with localized excited states in
  the energy gap},}\ }\href {\doibase 10.1007/BF01318313} {\bibfield  {journal}
  {\bibinfo  {journal} {Zeitschrift f{\"u}r Physik B Condensed Matter}\
  }\textbf {\bibinfo {volume} {47}},\ \bibinfo {pages} {217--222} (\bibinfo
  {year} {1982})}\BibitemShut {NoStop}%
\bibitem [{\citenamefont {Chubukov}\ \emph {et~al.}(2020)\citenamefont
  {Chubukov}, \citenamefont {Abanov}, \citenamefont {Esterlis},\ and\
  \citenamefont {Kivelson}}]{Chubukov2020}%
  \BibitemOpen
  \bibfield  {author} {\bibinfo {author} {\bibfnamefont {Andrey~V.}\
  \bibnamefont {Chubukov}}, \bibinfo {author} {\bibfnamefont {Artem}\
  \bibnamefont {Abanov}}, \bibinfo {author} {\bibfnamefont {Ilya}\ \bibnamefont
  {Esterlis}}, \ and\ \bibinfo {author} {\bibfnamefont {Steven~A.}\
  \bibnamefont {Kivelson}},\ }\bibfield  {title} {\enquote {\bibinfo {title}
  {Eliashberg theory of phonon-mediated superconductivity — when it is valid
  and how it breaks down},}\ }\href {\doibase
  https://doi.org/10.1016/j.aop.2020.168190} {\bibfield  {journal} {\bibinfo
  {journal} {Annals of Physics}\ }\textbf {\bibinfo {volume} {417}},\ \bibinfo
  {pages} {168190} (\bibinfo {year} {2020})},\ \bibinfo {note} {Eliashberg
  theory at 60: Strong-coupling superconductivity and beyond}\BibitemShut
  {NoStop}%
\bibitem [{\citenamefont {Yuzbashyan}\ and\ \citenamefont
  {Altshuler}(2022{\natexlab{a}})}]{EmilEli1}%
  \BibitemOpen
  \bibfield  {author} {\bibinfo {author} {\bibfnamefont {Emil~A.}\ \bibnamefont
  {Yuzbashyan}}\ and\ \bibinfo {author} {\bibfnamefont {Boris~L.}\ \bibnamefont
  {Altshuler}},\ }\bibfield  {title} {\enquote {\bibinfo {title}
  {Migdal-Eliashberg theory as a classical spin chain},}\ }\href {\doibase
  10.1103/PhysRevB.106.014512} {\bibfield  {journal} {\bibinfo  {journal}
  {Phys. Rev. B}\ }\textbf {\bibinfo {volume} {106}},\ \bibinfo {pages}
  {014512} (\bibinfo {year} {2022}{\natexlab{a}})}\BibitemShut {NoStop}%
\bibitem [{\citenamefont {Yuzbashyan}\ and\ \citenamefont
  {Altshuler}(2022{\natexlab{b}})}]{EmilEli2}%
  \BibitemOpen
  \bibfield  {author} {\bibinfo {author} {\bibfnamefont {Emil~A.}\ \bibnamefont
  {Yuzbashyan}}\ and\ \bibinfo {author} {\bibfnamefont {Boris~L.}\ \bibnamefont
  {Altshuler}},\ }\bibfield  {title} {\enquote {\bibinfo {title} {Breakdown of
  the Migdal-Eliashberg theory and a theory of lattice-fermionic
  superfluidity},}\ }\href {\doibase 10.1103/PhysRevB.106.054518} {\bibfield
  {journal} {\bibinfo  {journal} {Phys. Rev. B}\ }\textbf {\bibinfo {volume}
  {106}},\ \bibinfo {pages} {054518} (\bibinfo {year}
  {2022}{\natexlab{b}})}\BibitemShut {NoStop}%
\bibitem [{\citenamefont {Yuzbashyan}\ \emph {et~al.}(2024)\citenamefont
  {Yuzbashyan}, \citenamefont {Altshuler},\ and\ \citenamefont
  {Patra}}]{Emil2024}%
  \BibitemOpen
  \bibfield  {author} {\bibinfo {author} {\bibfnamefont {Emil~A.}\ \bibnamefont
  {Yuzbashyan}}, \bibinfo {author} {\bibfnamefont {Boris~L.}\ \bibnamefont
  {Altshuler}}, \ and\ \bibinfo {author} {\bibfnamefont {Aniket}\ \bibnamefont
  {Patra}},\ }\href {https://arxiv.org/abs/2409.19562} {\enquote {\bibinfo
  {title} {Instability of metals with respect to strong electron-phonon
  interaction},}\ } (\bibinfo {year} {2024}),\ \Eprint
  {http://arxiv.org/abs/2409.19562} {arXiv:2409.19562 [cond-mat.str-el]}
  \BibitemShut {NoStop}%
\bibitem [{\citenamefont {Abrikosov}\ \emph {et~al.}(1963)\citenamefont
  {Abrikosov}, \citenamefont {Gorkov},\ and\ \citenamefont
  {Dzyaloshinski}}]{AGD}%
  \BibitemOpen
  \bibfield  {author} {\bibinfo {author} {\bibfnamefont {A.~A.}\ \bibnamefont
  {Abrikosov}}, \bibinfo {author} {\bibfnamefont {L.~P.}\ \bibnamefont
  {Gorkov}}, \ and\ \bibinfo {author} {\bibfnamefont {I.~E.}\ \bibnamefont
  {Dzyaloshinski}},\ }\href@noop {} {\emph {\bibinfo {title} {Methods of
  Quantum Field Theory in Statistical Physics}}}\ (\bibinfo  {publisher} {Dover
  Publications},\ \bibinfo {year} {1963})\BibitemShut {NoStop}%
\bibitem [{\citenamefont {Kamenev}(2011)}]{Kamenev2011}%
  \BibitemOpen
  \bibfield  {author} {\bibinfo {author} {\bibfnamefont {Alex}\ \bibnamefont
  {Kamenev}},\ }\href@noop {} {\emph {\bibinfo {title} {Field Theory of
  Non-Equilibrium Systems}}}\ (\bibinfo  {publisher} {Cambridge University
  Press},\ \bibinfo {year} {2011})\BibitemShut {NoStop}%
\end{thebibliography}


\end{document}